\documentclass[a4paper,10pt]{article}
\usepackage{tikz}
\usepackage{graphicx}

\usepackage{mathtools}
\usepackage{bm}
\usepackage{dsfont}
\usepackage{pifont}
\usepackage{braket}
\usepackage{xcolor}
\usepackage{authblk}

\usepackage[normalem]{ulem}
\usepackage[utf8]{inputenc}
\usepackage[english]{babel}
\usepackage{amsmath,amsfonts,amssymb}
\usepackage[scaled]{helvet}
\usepackage{url}
\usepackage[colorlinks=true, allcolors=blue]{hyperref}
\usepackage{xifthen}
\usepackage{textcomp} 
\usepackage{graphicx,xcolor}
\usepackage{wrapfig} 
\usepackage{colortbl}
\usepackage{booktabs}
\usepackage{algorithm}
\usepackage[noend]{algpseudocode}
\usepackage{changepage}
\usepackage[numbers,sort&compress]{natbib}
\usepackage{fancyhdr}  
\usepackage{lastpage}  
\usepackage[explicit]{titlesec}
\usepackage{enumitem} 
\usepackage[resetlabels]{multibib}
\usepackage{letltxmacro}
\usepackage{xparse}
\usepackage{silence}
\usepackage{xpatch}
\usepackage{titlesec}

\titleformat{\paragraph}[runin]{\bfseries}{\theparagraph}{1em}{}

\def\flagdraftfig{false}

\newcommand*\circled[1]{\tikz[baseline=(char.base)]{
            \node[shape=circle,draw,line width=0.75pt,inner sep=0.5pt] (char) {\bf\textsf{#1}};}}

\makeatletter
\newcommand*{\encircled}[1]{\relax\ifmmode\mathpalette\@encircled@math{#1}\else\@encircled{#1}\fi}
\newcommand*{\@encircled@math}[2]{\@encircled{$\m@th\bf\sf#1#2$}}
\newcommand*{\@encircled}[1]{%
  \tikz[baseline,anchor=base]{\node[draw,circle,outer sep=0pt,inner sep=.2ex] {\bf\textsf{#1}};}}
\makeatother

\definecolor{mycolpurple}{HTML}{6600a6}
\definecolor{mycolgreen}{HTML}{00a60d}
\definecolor{mycolblue}{HTML}{0058a6}
\definecolor{mycolorange}{rgb}{0.8, 0.5, 0.2}
\definecolor{mycolblack}{rgb}{0,0,0}

\def\C{\mathbb C}

\def\muT{\bm{\mu}}
\def\epT{\bm{\varepsilon}}
\def\ep{\varepsilon}

\def\IOmj{\ensuremath{\mathsf{I}_{\Omega_j}}}

\def\z{\boldsymbol{\hat{z}}}
\def\E{\boldsymbol{E}}
\def\dfrac{\displaystyle\frac}

\def\tensmur{\boldsymbol{\mu}_r}
\def\tensepsr{\boldsymbol{\varepsilon}_{r}}
\def\tensepsrzz{\boldsymbol{\varepsilon}_{r_{zz}}}
\def\dfrac{\displaystyle\frac}
\def\epsr{\ensuremath{\varepsilon_r}}

\def\curl{\ensuremath{\mathbf{curl}}}
\def\div{\mathrm{div}}
\def\grad{\ensuremath{\mathbf{grad}}}

\def\se{\ensuremath{e}}
\def\redom{{\color{red}\omega}}
\def\omsrc{{\color{mycolgreen}\omega_s}}
\def\omeig{{\color{black}\widetilde{\omega}}}
\def\om{\omega} 
\def\bn{\ensuremath{\mathbf{n}}}
\def\br{\ensuremath{\mathbf{r}}}
\def\bx{\ensuremath{\mathbf{x}}}
\def\by{\ensuremath{\mathbf{y}}}

\def\bk{\ensuremath{\mathbf{k}}}
\def\x{\ensuremath{\mathbf{x}}}

\newcommand{\Ompml}[1]{\Omega^ {{\scalebox{.4}{{PML}}}}_{#1}}

\newcommand{\Galerkind}[3]{\ensuremath{\displaystyle\int_{#3}#1\cdot #2\,\mathrm{d}{\Omega}}}
\newcommand{\Galerkindlin}[3]{\ensuremath{\displaystyle\int_{#3}#1\cdot #2\,\mathrm{d}{#3}}}
\newcommand{\Galerkinscal}[3]{\ensuremath{\displaystyle\int_{#3}#1\,\overline{#2}\,\mathrm{d}{\Omega}}}
\newcommand{\Galerkinscald}[3]{\ensuremath{\displaystyle\int_{#3}#1\,#2\,\mathrm{d}{\Omega}}}
\newcommand{\Galerkinscalline}[3]{\ensuremath{\displaystyle\int_{#3}#1\,#2\,\mathrm{d}{#3}}}

\definecolor{colorgui}{RGB}{244, 110, 66}
\definecolor{lightblue}{RGB}{219, 235, 252}
\definecolor{lightgreen}{RGB}{219, 252, 232}
\definecolor{lightorange}{RGB}{252, 237, 219}
\definecolor{lightpurp}{RGB}{252, 219, 219}
\definecolor{lightred}{RGB}{240, 219, 252}
\definecolor{lightyellow}{RGB}{255, 255, 230}
\definecolor{limegreen}{rgb}{0.2, 0.8, 0.2}
\definecolor{lincolngreen}{rgb}{0.11, 0.35, 0.02}

\def\tdm{truncation-dependent modes}

\title{Dispersive Perfectly Matched Layer and high order Absorbing Boundary Conditions for the computation of Quasinormal modes of open electromagnetic structures}
\author[1,*]{Guillaume Dem{\'e}sy}
\author[2]{Tong Wu}
\author[3]{Yoann Br{\^u}l{\'e}}
\author[1]{Fr\'ed\'eric Zolla}
\author[1]{Andr\'e Nicolet}
\author[2]{Philippe Lalanne}
\author[1]{Boris Gralak}

\affil[1]{Aix Marseille Univ, CNRS, Centrale Marseille, Institut Fresnel, Marseille, France}
\affil[2]{Université de Bordeaux, Institut d’Optique Graduate School, CNRS, LP2N, F-33400 Talence, France}
\affil[3]{ICB, Université de Bourgogne-Franche Comté, CNRS, Dijon, France}
\affil[*]{Corresponding author : \texttt{guillaume.demesy@fresnel.fr}.}

\def\C{\mathbb C}

\begin{document}
\maketitle

\begin{abstract}
	Resonances, also known as quasinormal modes (QNM) in the non-Hermitian case, play a ubiquitous role in all domains of physics ruled by wave phenomena, notably in continuum mechanics, acoustics, electrodynamics, and quantum theory. The non-Hermiticity arises from the system losses, whether they are material (Joule losses in electromagnetism) or linked to the openness of the problem (radiation losses). In this paper, we focus on the latter delicate matter when considering bounded computational domains mandatory when using \textit{e.g.} Finite Elements. Dispersive perfectly matched layers and absorbing boundary conditions are studied.
\end{abstract}

\section{Introduction}
An open structure described by its geometry, material properties and boundary conditions has a response to a solicitation which is strongly correlated to its intrinsic properties, and it appears that the strongest responses are related to its \textbf{resonances}, or quasinormal modes (QNM), the solutions of the wave equation without any source. Determining these eigenmodes is therefore extremely useful for both physical understanding and practical computations. The QNM formalism applies to the wave equation irrespective of the considered physics : quantum mechanics, relativity, acoustics \cite{koch2005acoustic}, elastodynamics \cite{PhysRevB.107.144301}, electromagnetism \cite{leung1994completeness}... For comprehensive reviews about QNMs in electromagnetism, which is the scope of this paper, the reader can refer to Refs.~\cite{lalanne2018light,kristensen2020modeling,sauvan2022normalization,wu2021nanoscale,both2021resonant}.

A typically realistic optically resonant structure consists of scatterers often made of lossy dispersive materials (semi-conductors or noble metals, exhibiting Joule losses) merged into media of spatial extension that can be considered as being infinite in view of the operating wavelength of the device. Taking into account the frequency-dependence of materials in eigenvalue problems turns them into \textbf{non-linear eigenvalue problems} which is obvious writing the source free Helmholtz equation as : $c^2\tensepsr(\bx,\redom)^{-1}\curl\,[\tensmur^{-1}(\bx)\curl\,\E]=\redom^2\E$, where $\tensepsr(\bx,\redom)$ is the dispersive relative permittivity. The \textit{theoretical} spectrum of such a lossy structure, that is, its set of eigenvalues, is complex in two ways since the eigenvalues belong to the complex plane and since their location in the complex plane is far from being trivial. The spectrum features an infinite set of isolated eigenvalues whose eigenmodes are the QNMs, accumulation points linked to materials intrinsic resonances, \textit{e.g.} the poles of a Lorentz permittivity model, essential spectrum linked to the presence of corners \cite{demesy2020non}. In turn, their \textit{numerical} determination is a difficult task. Given the high stakes of the modal point of view, these linear (in the non-dispersion case) or non-linear (dispersive case) eigenvalue problems and associated modal expansion formalism have received a lot of attention in photonics \cite{leung1994completeness,bai2013efficient,sauvan2013theory,vial2014quasimodal,muljarov2016resonant,perrin2016eigen,yan2018rigorous,zolla2018photonics,Lalanne:19,BinkowskiQNM2020,BinkowskiRiesz2020,campos2020computation,truong2020continuous,demesy2020non} for the last decade. 

In academic cases where the compact resonator lies in a uniform medium (no substrate), QNM expansion leads to a convergent series inside the resonator, as thoroughly shown and numerically exploited in Refs.~\cite{zschiedrich2018riesz,lobanov2019resonant,sehmi2020applying,gras2020nonuniqueness,sauvan2022normalization,wu2023modal,besbes2022role,yan2018rigorous,sauvan2021quasinormal}. This allows to use QNM to perform modal expansion, \textit{i.e.} to deduce the response of the system to a given source at a frequency $\redom_s$ as a sum over QNMs weighted by coupling coefficients deduced by mere overlap integrals between the source and the QNMs. Even in more realistic cases were completeness cannot be proven, modal expansion gives good results \cite{yan2018rigorous,gras2019quasinormal}. 

In this paper, our focus is on the unbounded or open nature of the problem is another source of losses, now due to radiation at infinity. Again, the problem becomes non-Hermitian, which can be seen formally at the continuous level when writing the outgoing wave conditions at infinity, namely the Sommerfeld (or Silver-M{\"u}ller) \textbf{radiation condition} as: $\lim_{|\br|\rightarrow\infty}|\br|\,[\curl\,\E-i\redom\,\br\times\E/(|\br|c)]=0$. 

Popular methods traditionally used in electromagnetism for direct problems have been adapted for calculating QNMs : vector spherical harmonics \cite{STOUT2018173,PhysRevB.98.085418}, Fourier Modal Method \cite{sauvan2013theory,Weiss:11}, volume \cite{Kristensen:12} and surface (boundary elements) \cite{alpeggiani2016visible} integral methods, Finite Difference Frequency Domain method (FDFD) \cite{zimmerling2016lanczos,zimmerling2016efficient}, Finite Element Method (FEM) \cite{vial2014quasimodal}, etc. The first four cited methods do not require to mesh the infinite background and the outgoing wave condition is enforced through a proper choice of basis function \textit{i.e.} outgoing spherical harmonic, Fourier order, or Green function respectively. Note that the eigenfrequency appears explicitly in the basis vector of those methods, and cannot be factorized outside the corresponding matrices. As a consequence, the eigenvalue problem is solved using contour integration \cite{beyn2012integral}. The two latter fully grid-based methods FDFD and FEM rely on a mesh of the whole structure, including freespace away from the resonator and require to define a proper outgoing wave condition at a finite distance from the resonator. The discretization in these two methods is purely space-based and the eigenfrequency can be factorized from the matrices, allowing to use specific algorithms based on Krylov subspaces from the SLEPc library \cite{Hernandez:2005:SSF} for instance.

A very common way to deal with unbounded wave problems with a finite computational domain, in order to emulate the Sommerfeld condition taking place at infinity on the border of the computational domain of interest, is to introduce an absorbing condition at finite distance involving complex coefficients. The imaginary part of these coefficients is taking into account the outgoing power never coming back. Two vast families of methods can be invoked, a volume one and a surface one.
\begin{itemize}
  \item The first one is to use a \textbf{Perfectly Matched Layer} (PML) which was first introduced in the time domain as a dispersive object \cite{pml_berenger,chew19943d} since the damping is expected to be the same for all the frequencies forming the numerical pulse. So far, the PMLs used in QNM computation have been chosen to be non-dispersive, choosing their parameters to target a specific region of the complex spectrum. Note that there has been recent attempts to implement dispersive PMLs, in cylindrical \cite{nannen2018computing} and Cartesian coordinates \cite{gras2020nonuniqueness}. 
  \item The second one is to adapt the Sommerfeld condition at finite distance leading to a so called Dirichlet-to-Neumann map linking the Dirichlet data, related to $\E$, to Neumann ones, related to $\curl\,\E$. These surface condition are referred to as \textbf{Absorbing Boundary Conditions} (ABC). Note that the simplest version of the ABC, called first order ABC or Mur ABC, has been used in the context of QNMs computations \cite{nanostructured2022hessler}. A general modal boundary condition can also be found in \cite{ARAUJOC2021110024}.
\end{itemize}

Applying these truncation techniques to model infinite extents splits the modes into two categories: the PML modes or Truncation-Dependent modes (TDMs) that depend on the PML or ABC parametrization, \textit{e.g.} the distance between the object and the truncation boundary, and those which do not (QNMs) \cite{vial2014quasimodal,lalanne2018light,yan2018rigorous}. The PML modes can be further classified into two types \cite{yan2018rigorous}. The first type of PML modes is formed by waves that are bouncing back and forth between the PML boundaries. They dominantly correspond to the discretization of the continuous spectrum \cite{vial2014quasimodal}. The second type of PML modes originate from QNMs, of which outgoing waves are not sufficiently damped and reflected back. It is well known that both types of PML modes must be considered to guarantee the completeness of the modal expansion approach. 
The fact that the modal expansion depends on a large number of PML modes or TDMs is a significant impediment to the wide use of the method. 
Numerically, computing PML modes is challenging. The field distribution of PML modes in the high-frequency and low-Q region varies significantly, requiring an ultrafine mesh (and/or a high FEM order) within the PML domain for accurate computation. Moreover, PML modes are numerous and spread throughout the complex plane. Finding all the necessary PML modes for the convergence of the modal expansion is time-consuming and computationally expensive.
Physically, individual PML modes are devoid of physical meaning, and considering a large number of PML modes can significantly dilute the physical insights that the model expansion can offer. Certain PML modes, particularly those originating from QNMs as mentioned earlier, may contribute significantly to the main features of the spectrum (peaks or dips), making it challenging to interpret the physics using the modal approach.
The question is whether there is a specific type of PML or ABC that can mitigate the disadvantages mentioned above. Ideally, all the PML modes and QNMs can be easily and accurately computed; the PML modes and QNMs are disentangled (no PML mode originating from QNMs); and the number of PMLs determining the response is kept at a reasonable number. In most previous modal expansion works, non-dispersive PMLs have been used \cite{sauvan2022normalization,demesy2020non,zolla2018photonics,vial2014quasimodal,yan2018rigorous,truong2020continuous,gras2020nonuniqueness,wu2023modal,gras2019quasinormal,besbes2022role}. This at least causes two problems: first, in the low-frequency range, the damping factor is too small, and a large number of low-frequency QNMs are contaminated: they are not “unveiled” and degrade into PML-like modes \cite{vial2014quasimodal,sauvan2022normalization}; second, in the high-frequency range, the field inside the PML layer decays too fast, such that an accurate computation of the modes becomes difficult, and a large number of spurious modes emerge \cite{trefethen2006nonhermitian,ARAUJOC2021110024}. These drawbacks may be overcome by implementing dispersive PMLs or ABCs. The former offers an absorption coefficient that is inversely proportional to the frequency, which allows the decay rate inside the PML to be equal for all frequencies. The later replaces the PML domain with a zero-thickness boundary condition, which can possibly reduce the number of TDMs.

\color{mycolblack}
The paper is organized as follows. In a first part (Sec.~\ref{sec:PML}), we study dispersive PMLs. The 1D case is extensively treated and exhibit interesting properties, notably on the location in the complex plane of eigenfrequenies  of the \tdm{}. Then, we move to the 2D scalar case and show the implementation of dispersive PMLs in order to damp the electromagnetic field at the same rate for all the computed QNMs. We discuss why the 1D ideal case is difficult to extend to the 2D scalar case with Cartesian PMLs with both constant or graded absorption profiles. In a second part (Sec.~\ref{sec:dtn}), we focus on another type of outgoing wave condition, the ABC, which is a surface one. We compare the QNMs obtained with PMLs to those obtained using a local approximation of the Dirichlet-to-Neumann operator of the first or second order instead of PMLs, which brings an original angle to the discretization of the continuous spectrum.

\section{Dispersive Perfectly Matched Layers}\label{sec:PML}

Perfectly matched layers (PML) are numerical and theoretical tools allowing to handle the oscillating outgoing nature of waves in unbounded problems. In direct frequency domain, the wavelength in the infinite space to truncate is fixed which facilitates the optimization of the PML parameters (damping profile, truncation, mesh discretization...) In spectral problems, PML are a widely used theoretical tool allowing to retrieve the quasinormal modes. However, in the modal context, the wavelength becomes eigenvalue dependent and the use of non-dispersive PML, well optimized for direct harmonic problems where the wavelength is fixed, leads to a radically different treatment of low and high frequency QNMs. Thus, we propose to investigate the use of frequency-dependent PML damping profiles, as it was originally the case in the seminal paper by Bérenger which was targeting time-domain methods (FDTD). Indeed, in time domain methods, the source is a short pulse containing a broad range of frequencies whose individual decay in the PML ought to be handled similarly to prevent reflection of the lower frequencies at the end of PML. Note that this approach has already been studied in Ref.~\cite{nannen2018computing} in the case of cylindrical dispersive PMLs. We investigate Cartesian PMLs as they are quite suitable to the multiple layers and substrate often encountered in optics.


In this section, we first study analytically the behavior of traditional dispersion-free PML and dispersive PML on the classical 1D slab problem. Then, we show a possible Finite Element implementation relying on a polynomial eigenvalue problem.

\subsection{A 1D toy model}
\begin{figure}[ht]
\centering
\begin{tikzpicture}[scale=1]
\def \a{2};
\draw [<-] (-2.5\a+0*\a,\a/2) node[left]{$y$} -- (-2.5\a+0*\a,0);
\draw [->] (-2.5\a+0*\a,0) circle (4 pt) node[left]{$z$} -- (-2.5\a+6*\a,0);
\draw [->] (-2.5\a+0*\a,0) -- (-2.5\a+6*\a,0) node[below]{$x$};
\filldraw
  (-2.5\a+0*\a,0) circle (2pt) node[align=center,above] {$\Gamma_0$} -- 
(-2.5\a+1*\a/2,0) node[align=center,below] {$\Omega_1$\\PML} --
  (-2.5\a+1*\a,0) circle (2pt) node[align=center,above] {$\Gamma_1$} -- 
(-2.5\a+3*\a/2,0) node[align=center,below] {$\Omega_2$\\air} --
  (-2.5\a+2*\a,0) circle (2pt) node[align=center,above] {$\Gamma_2$} -- 
(-2.5\a+5*\a/2,0) node[align=center,below] {$\Omega_3$\\slab} --
  (-2.5\a+3*\a,0) circle (2pt) node[align=center,above] {$\Gamma_3$} -- 
(-2.5\a+7*\a/2,0) node[align=center,below] {$\Omega_4$\\air} --
  (-2.5\a+4*\a,0) circle (2pt) node[align=center,above] {$\Gamma_4$} -- 
(-2.5\a+9*\a/2,0) node[align=center,below] {$\Omega_5$\\PML} --
  (-2.5\a+5*\a,0) circle (2pt) node[align=center,above] {$\Gamma_5$};
\end{tikzpicture}
\caption{Sketch and notations of the 1D toy model.}
\label{fig:1Dsketch}
\end{figure}
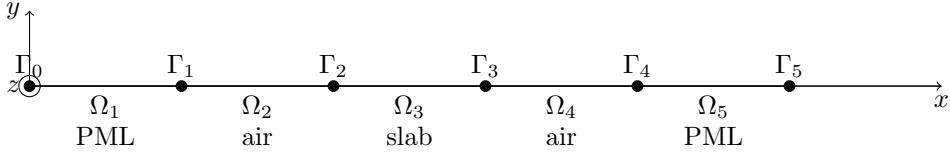

We first consider a mere dielectric slab. The structure equipped with PMLs is made of five linear and homogeneous layers $\Omega_j$, infinite along both axes  $O_z$ and  $O_y$, with relative permittivities $\epT_{r,j}$ and relative permeabilities $\muT_{r,j}$ with $j\in J = \{\,1,\,2,\,3,\,4,\,5\}$ (see Fig.~\ref{fig:1Dsketch}). The thickness of each layer is denoted $a_j$.\\

Domains $\Omega_1$ et $\Omega_5$ are two PMLs absorbing along the $x$ direction. The relative permittivity and permeability tensors have the following expressions:
\begin{equation}
\epT_{r,1} = \muT_{r,1} = \begin{bmatrix}
1/\eta_1&0&0\\
0&\eta_1&0\\
0&0&\eta_1
\end{bmatrix} \mbox{ and }
\epT_{r,5} = \muT_{r,5} = \begin{bmatrix}
1/\eta_5&0&0\\
0&\eta_5&0\\
0&0&\eta_5
\end{bmatrix}\, ,
\end{equation}
where $\eta_1$ is characterized the complex stretch of the PML, which can be a mere complex number independent of the frequency ($\eta_1 = \alpha_1 + i\beta_1$ with $\{\alpha_1,\beta_1\} \in\mathbb{R}_+^2$) or frequency dependent in the case of dispersive PMLs ($\eta_1 = (\alpha_1 + i\beta_1)/\redom$). The same holds for the other PML domain $\Omega_5$ with $\eta_5 = \alpha_5+i\beta_5$.

The domains $\Omega_2$ and $\Omega_4$ are made of freespace, \textit{i.e.} $\ep_{r,2}=\ep_{r,4}=\mu_{r,2}=\mu_{r,4} = \mathds{1}$. Finally, the central domain  $\Omega_3$  is made of an amagnetic homogeneous material with relative permittivity $\epT_{r,3}=\ep_{r,3}\mathds{1}$. We introduce the characteristic functions $\IOmj(\mathbf{r})=1$ for $\mathbf{r}\in\Omega_j$ and $O$ elsewhere. Finally, the permittivity tensor field of the total structure is defined as:
\begin{equation}
\epT_{r} = \sum_j \epT_{r,j}\,\IOmj(\mathbf{r}) = \sum_j
\begin{bmatrix}
\ep_{x,j}&0&0\\
0&\ep_{y,j}&0\\
0&0&\ep_{z,j}
\end{bmatrix}\IOmj(\mathbf{r})
\end{equation}
where $\ep_{x,j},\,\ep_{y,j}$ et $\ep_{z,j}$ are the diagonal components of each tensor $\epT_{r,j}$ and the same notations hold for the relative permeability tensor field of the total structure $\muT_r$.

The two interface of each domain $\Omega_j$ are denoted $\Gamma_{j-1}$ and $\Gamma_j$ as shown in Fig.~\ref{fig:1Dsketch}. Finally, let us define the coordinates $b_k$ of each interface $\Gamma_{k}$ as $b_k=\sum_{j=1}^k a_j$ for $k\in\{1\dots5\}$ and $b_0=0$. Dirichlet boundary condition are set on $\Gamma_0$ and $\Gamma_5$.

\begin{figure}[h!]
  \centering
  \includegraphics[draft=\flagdraftfig,width=0.98\textwidth]{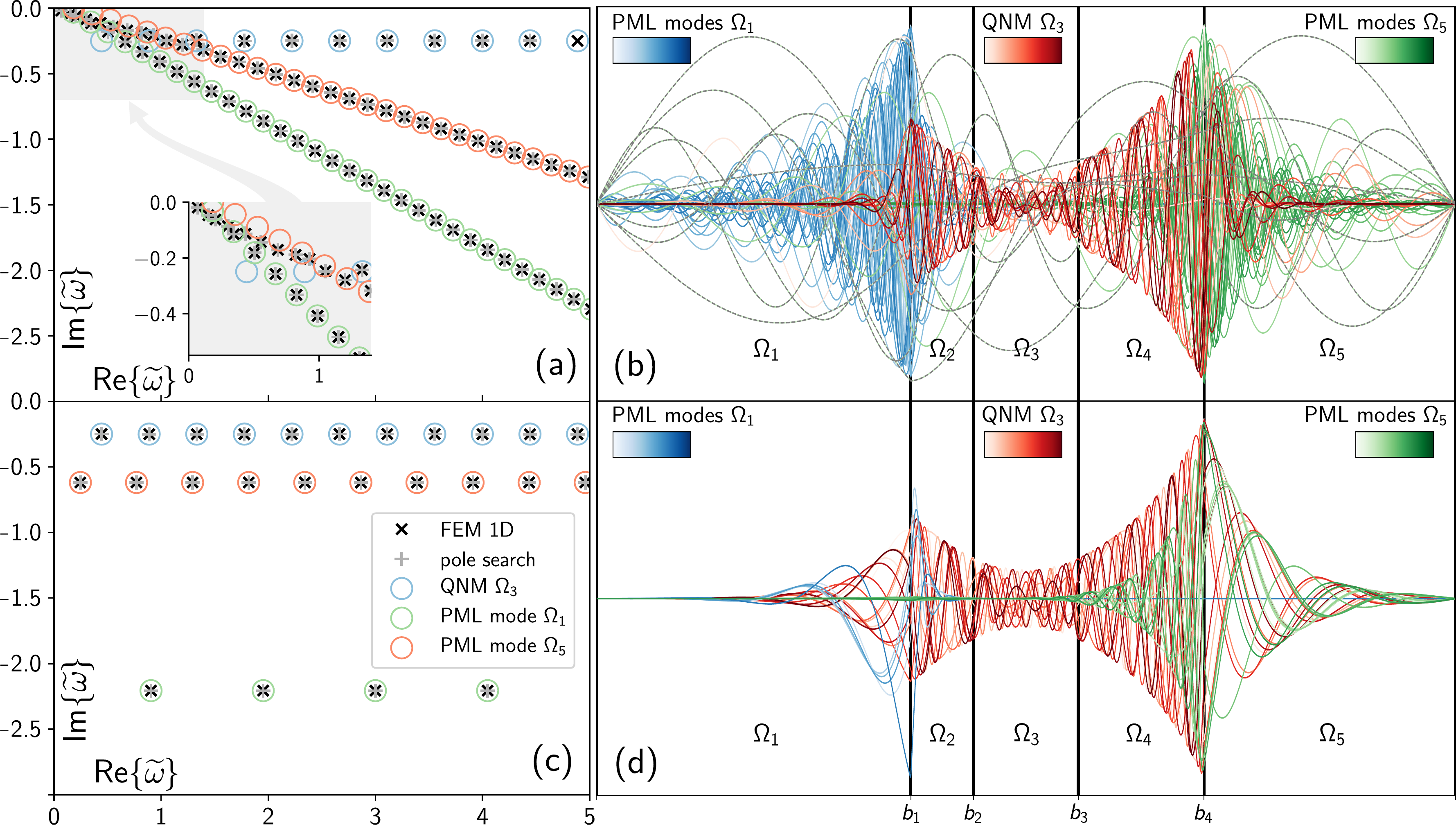}
  \caption{Eigenfrequencies (a) and eigenmodes (b) of the slab for non-dispersive PMLs. Eigenfrequencies (c) and eigenmodes (d) of the slab for dispersive PMLs. In (a) and (c), black crosses correspond to the FEM results, gray pluses to the pole search results and colored circles to the asymptotic approach. }
  \label{fig:oned}
\end{figure}

We choose to deal with the electric field $\E(\mathbf{r},\om)$ which reduced to the scalar unknown function $e(x,\om)$ such that $\E(\mathbf{r},\om) = e(x,\om)\z$. The spectral problem amounts to look for non-trivial solutions $(\redom,e)\in\C\times \bm{\mathcal{H}}_0^1(\Omega)$ of:
\begin{equation}\label{eq:oned}
  \left\{
  \begin{array}{rcll}
    -\mu_{y,j}^{-1}\,e''(x,\redom) &=& \ep_{z,j}\dfrac{\redom^2}{c^2} \text{ in } \Omega_j & \mbox{ with } j \in \{1,\,2,\,3,\,4,\,5\} \\[2mm]
    \mu_{y,j}^{-1}\, e'(b_j^-,\redom) &=& \mu_{y,(j+1)}^{-1} e'(b_j^+,\redom) & \mbox{ with } j \in \{1,\,2,\,3,\,4\}  \\[2mm]
    e(b_j^-,\redom) &=& e(b_j^+,\redom) & \mbox{ with } j \in \{1,\,2,\,3,\,4\}\\[2mm]
    e(0,\redom) &=& e(b_5,\redom) = 0&  \\
  \end{array}\right. .
\end{equation}
We conveniently set $c=1$ and let $x_1(\redom) = \redom(a_1\eta_1+a_2)$, $x_5(\redom) = \redom(a_5\eta_5+a_4)$ and $x_3(\redom) = \redom a_3\sqrt{\epsilon_3}$. Applying the classical transfer matrix approach leads to the following transcendental equation fulfilled by the modes of the structure:
\begin{equation}
\begin{array}{ll}
  &\dfrac{\cos[x_3(\redom)]}{\redom}\left(\cos[x_1(\redom)]\sin[x_5(\redom)]+\sin[x_1(\redom)]\cos[x_5(\redom)]\right)\\[3mm]
  +&\dfrac{\sin[x_3(\redom)]}{\redom}\left(\dfrac{1}{\sqrt{\epsilon_3}}\cos[x_1(\redom)]\cos[x_5(\redom)] - \sqrt{\epsilon_3}\sin[x_1(\redom)]\sin[x_5(\redom)]\right)=0\, .\\
\end{array}
\label{eq:trans}
\end{equation}
In the following, we have solved Eq.~\ref{eq:oned} in three ways for both dispersive and non-dispersive PMLs: (i) using a brute Finite Element discretization, (ii) via a numerical pole search approach of Eq.~\ref{eq:trans}, and (iii) asymptotically as detailed hereafter. The corresponding eigenfrequencies are shown in the complex frequency plane in Fig.~\ref{fig:oned}(a) for the non-dispersive PML case (respectively in Fig.~\ref{fig:oned}(c) for the dispersive PML case) in black crosses for the FEM (i), gray pluses for the pole search (ii) and colored circles for the asymptotic approach (iii) in the non-dispersive (respectively dispersive) PML case. The values used in these numerical experiments are the following: $\{a_1,a_2,a_3,a_4,a_5\}=\{15,3,5,6,12\}$, $\eta_1 = e^{i\pi/6}$, $\eta_5 = e^{i\pi/8}$, $\ep_{r,3}=2$.

\subsubsection{QNMs of the slab $\Omega_3$}
A first asymptotic study allows to determine the so called of the system QNM. Let $\eta_1 = \eta_5=1$ and $a_1=a_5\rightarrow\infty$. After derivations, one can retrieve for the asymptotic expressions of the eigenfrequencies $\omeig$:

\begin{equation}\label{eq:QNM}
\omeig  = \dfrac{-i}{\sqrt{\epsilon_3} a_3}\ln\left[\left\lvert r \right\lvert\right] + \dfrac{k\pi}{\sqrt{\epsilon_3} a_3} \mbox{,  where } r = \dfrac{1-\sqrt{\epsilon_3}}{1+\sqrt{\epsilon_3}}
\end{equation}
is the reflection coefficient of the air/$\Omega_3$ interface and $k \in \mathbb{N}$. (Note that tilded quantities are used throughout the paper for analytically or numerically computed eigenvalues and eigenvectors.) These eigenvalues of Eq.~\ref{eq:oned} are the QNM of the slab $\Omega_3$.  They correspond to poles of the scattering matrix of a single layer of relative permittivity $\ep_3$ and thickness $a_3$. The slab supports an infinity of QNM with identical imaginary part proportional to the logarithm of the reflection coefficient of one interface air/material of permittivity $\ep_3$. Their real part are spaced by $\pi/(\sqrt{\epsilon_3} a_3)$. The complex frequencies corresponding to Eq.~\ref{eq:QNM} are shown in Fig.~\ref{fig:oned}(a) and (c) in blue circles. They are independent of the choice of the type and parameters of the PMLs. The corresponding eigenvectors are shown in red colors in Fig.~\ref{fig:oned}(b) and (d). As desired, QNM decay at the same rate in the PMLs in the dispersive-PML case.

\subsubsection{Truncation Dependent Modes (TDMs): PML modes of $\Omega_1$ et $\Omega_5$.}\label{PMLNODISPSEC}

Let us now focus on the TDMs, also called the PML modes, \textit{e.g.} modes of domain $\Omega_1$, and use another asymptotic approximation. First, the thickness of layer $\Omega_3$ is now considered infinite: $a_3\rightarrow\infty$. This approximation allows to artificially discard all the QNM.  This can be retrieved in Eq.~\ref{eq:QNM}, with $a_3\rightarrow\infty$, the QNM frequencies reduce to the sole null frequency.

Now, since we are focusing on PML modes of the domain $\Omega_1$, let us also artificially consider that $\Omega_5$ is a semi-infinite freespace domain by letting $a_5\rightarrow\infty$ and $\eta_1=1$. These assumptions now lead to:

\begin{equation}
\tan\left[\redom(a_1\eta_1+a_2)\right] = \dfrac{i}{\sqrt{\epsilon_3}}\,,
\label{PML5}
\end{equation}
and we obtain the eigenfrequencies associated with the PML $\Omega_1$:
\begin{equation}
\redom(a_1\eta_1+a_2) = \arctan\left[ \dfrac{i}{\sqrt{\epsilon_3}}\right] + k\pi
\label{PML6}
\end{equation}
with $k \in \mathbb{N}$.
Now two case should be considered, depending on whether the PML is considered dispersive or not.

\paragraph*{Non-dispersive PML.} In this case, the PML stretch parameter is given by $\eta_1 = \alpha_1 + i\beta_1$ and Eq.~\ref{PML6} becomes:
\begin{equation}
  \omeig = \left[\dfrac{-i\ln[|r|]}{2}+k\pi\right]\left[\dfrac{a_1\alpha_1+a_2-i a_1\beta_1}{(a_1\alpha_1+a_2)^2+a_1^2\beta_1^2}\right], \, k\in \mathbb{N}\, .
  \label{eq:noPMLdisp}
\end{equation}
From this last expression \ref{eq:noPMLdisp}, we deduce that there exist an infinity of PML modes for the (bounded) domain $\Omega_1$. In the complex plane, their eigenvalues are evenly spaces on a straight line with guiding coefficient $\dfrac{-a_1\beta_1}{a_1\alpha_1+a_2}$. It is also important to note that when $a_2 \ll a_1$, this coefficient reduces to $-\beta_1/\alpha_1$, and thus the continuous spectrum is rotated by an angle opposite to the argument of the PML stretch $\eta_1$. Indeed, this type of PML is efficient at high frequency even with a reasonable size $a_1$, but fails at damping the QNMs at lower frequencies. The complex frequencies corresponding to Eq.~\ref{eq:noPMLdisp} are shown in Fig.~\ref{fig:oned}(a) in red and green circles. Note that due to the PML finite thickness, the rotated continuous spectrum becomes discrete. 

\paragraph*{Dispersive PML.} Now, let $\eta_{1} = (\alpha_1 + i\beta_1)/\redom$  and Eq.~\ref{PML6} becomes:
\begin{equation}
\omeig = k\dfrac{\pi}{a_2} - \dfrac{a_1(\alpha_1 + i \beta_1)}{a_2} -\dfrac{i}{2 a_2}\ln\left[|r|\right] , \, k\in \mathbb{N}\, .
\label{eq:PMLdisp}
\end{equation}
This new expression of the eigenfrequencies of the dispersive PML modes  of $\Omega_1$ shows that there still exist an infinity of modes but now their eigenfrequencies lie on a \textit{straight line parallel to the real axis} with imaginary part $(-\ln\left[|r|\right]/2-a_1\beta_1)/a_2$. The complex frequencies corresponding to Eq.~\ref{eq:noPMLdisp} are shown in Fig.~\ref{fig:oned}(b) in red and green circles. Thus, contrarily to non-dispersive PMLs, the damping induced by this eigenvalue dependent stretch is now the same independently of the real part of the considered mode. We indeed retrieve the very important historical property of the time-domain PML where an ultra-short pulse are at stake. This discrete mode pattern on a horizontal line in the complex plane is no longer a \textit{rotation} of the real axis as it was the case with dispersion-free PMLs, but a \textit{translation} of the real axis.

\subsection{A naive extension to the 2D scalar case}\label{part:fem2Ddisppml}
\subsubsection{Setup and notations}
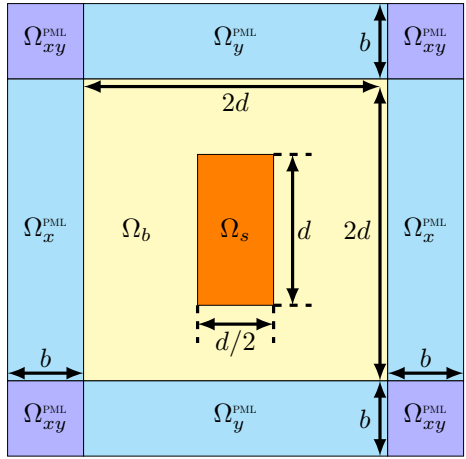
\begin{figure}[h]
  \centering
\begin{tikzpicture}
\draw[very thin, color=black, fill=blue!30] (-2,-2) rectangle (-3,-3);
\draw[very thin, color=black, fill=blue!30] (2,2) rectangle (3,3);
\draw[very thin, color=black, fill=blue!30] (-2,2) rectangle (-3,3);
\draw[very thin, color=black, fill=blue!30] (2,-2) rectangle (3,-3);
\draw[very thin, color=black, fill=cyan!30] (-2,-2) rectangle (2,-3);
\draw[very thin, color=black, fill=cyan!30] (-2,2) rectangle (2,3);
\draw[very thin, color=black, fill=cyan!30] (-2,2) rectangle (-3,-2);
\draw[very thin, color=black, fill=cyan!30] (2,2) rectangle (3,-2);
\draw[very thin, color=black, fill=yellow!30] (-2,-2) rectangle (2,2);
\draw[very thin, color=black, fill=orange] (-0.5,-1) rectangle (0.5,1);
\draw[very thick, color=black, style=dashed] (-0.5,-1) -- (-0.5,-1.5);
\draw[very thick, color=black, style=dashed] (0.5,-1) -- (0.5,-1.5);
\draw[very thick, color=black, <->, >=latex] (-0.5,-1.25) -- (0.5,-1.25);
\draw[very thick, color=black, style=dashed] (0.5,1) -- (1,1);
\draw[very thick, color=black, style=dashed] (0.5,-1) -- (1,-1);
\draw[very thick, color=black, <->, >=latex] (0.75,-1) -- (0.75,1);
\node at (0,-1.5) {$d/2$};
\node at (0.9,0) {$d$};
\draw[very thick, color=black, <->, >=latex] (1.9,-2) -- (1.9,1.9);
\node at (1.6,0) {$2d$};
\draw[very thick, color=black, <->, >=latex] (-2,1.9) -- (1.9,1.9);
\node at (0,1.7) {$2d$};
\draw[very thick, color=black, <->, >=latex] (1.9,3) -- (1.9,2);
\draw[very thick, color=black, <->, >=latex] (1.9,-3) -- (1.9,-2);
\node at (1.7,2.5) {$b$};
\node at (1.7,-2.5) {$b$};
\draw[very thick, color=black, <->, >=latex] (-2,-1.9) -- (-3,-1.9);
\draw[very thick, color=black, <->, >=latex] (2,-1.9) -- (3,-1.9);
\node at (2.5,-1.7) {$b$};
\node at (-2.5,-1.7) {$b$};
\node at (-1.3,0) {$\Omega_b$};
\node at (0,0) {$\Omega_s$};
\node at (2.5,0)    {$\Ompml{x}$};
\node at (-2.5,0)   {$\Ompml{x}$};
\node at (0,2.5)    {$\Ompml{y}$};
\node at (0,-2.5)   {$\Ompml{y}$};
\node at (2.5,2.5)  {$\Ompml{xy}$};
\node at (-2.5,2.5) {$\Ompml{xy}$};
\node at (-2.5,-2.5){$\Ompml{xy}$};
\node at (2.5,-2.5) {$\Ompml{xy}$};
\end{tikzpicture}
\caption{Two-dimensional rectangular resonator (orange region, dimensions $d/2$ by $d$) in freespace (light yellow region, dimensions $2d$ by $2d$) surrounded by Cartesian PMLs (blueish regions, width $b=1.5d$).}
\label{fig:twodpmls}
\end{figure}
 

In this section, the 2D scalar case is considered. As shown in Fig.~\ref{fig:twodpmls}, a $d/2$ by $d$ rectangular scatterer ($\tensepsr=4\mathds{1}$ in $\Omega_s$) lying in a portion of dimensions $2d$ by $2d$ of freespace background ($\tensepsr=\mathds{1}$ in $\Omega_b$) is surrounded by finite Cartesian PMLs of thickness $b=1.5d$. The relative permittivity and permeability tensors of dispersive PMLs are by $\tensmur^{\mathrm{PML}}(\bx,\redom)=\tensepsr^{\mathrm{PML}}(\bx,\redom)=\mathrm{Diag}[L_{xx}(\bx,\redom),L_{yy}(\bx,\redom),L_{zz}(\bx,\redom)]$ with $L_{xx}=s_y s_z/s_x$, $L_{yy}=s_x s_z/s_y$, $L_{zz}=s_x s_y/s_z$ and:
\begin{equation}
  \left\{
    \begin{array}{l}
      s_x(\bx,\redom) = \eta/\redom \mbox{ and }  s_y(\bx,\redom) = s_z(\bx,\redom) = 1            \mbox{ in } \Ompml{x},\\[2mm]
      s_y(\bx,\redom) = \eta/\redom \mbox{ and }  s_z(\bx,\redom) = s_x(\bx,\redom) = 1            \mbox{ in } \Ompml{y},\\[2mm]
      s_z(\bx,\redom) = 1           \mbox{ and }  s_x(\bx,\redom) = s_y(\bx,\redom) = \eta/\redom  \mbox{ in } \Ompml{xy},\\[2mm]
      s_x(\bx,\redom) = s_y(\bx,\redom) = s_z(\bx,\redom) = 1 \mbox{ in } {\Omega_b\cup\Omega_s}.
    \end{array}
    \label{eq:defpml}
  \right.
\end{equation}
The complex stretch $\eta$ in the above definition is set to $e^{i\pi/4}$. Finally, homogeneous Neumann boundary conditions are applied on the outer boundary of $\Omega:=\Omega_s\cup\Omega_b\cup\Ompml{x}\cup\Ompml{y}\cup\Ompml{xy}$.

\begin{figure}[h] 
  \centering 
  \includegraphics[draft=\flagdraftfig,width=.9\textwidth]{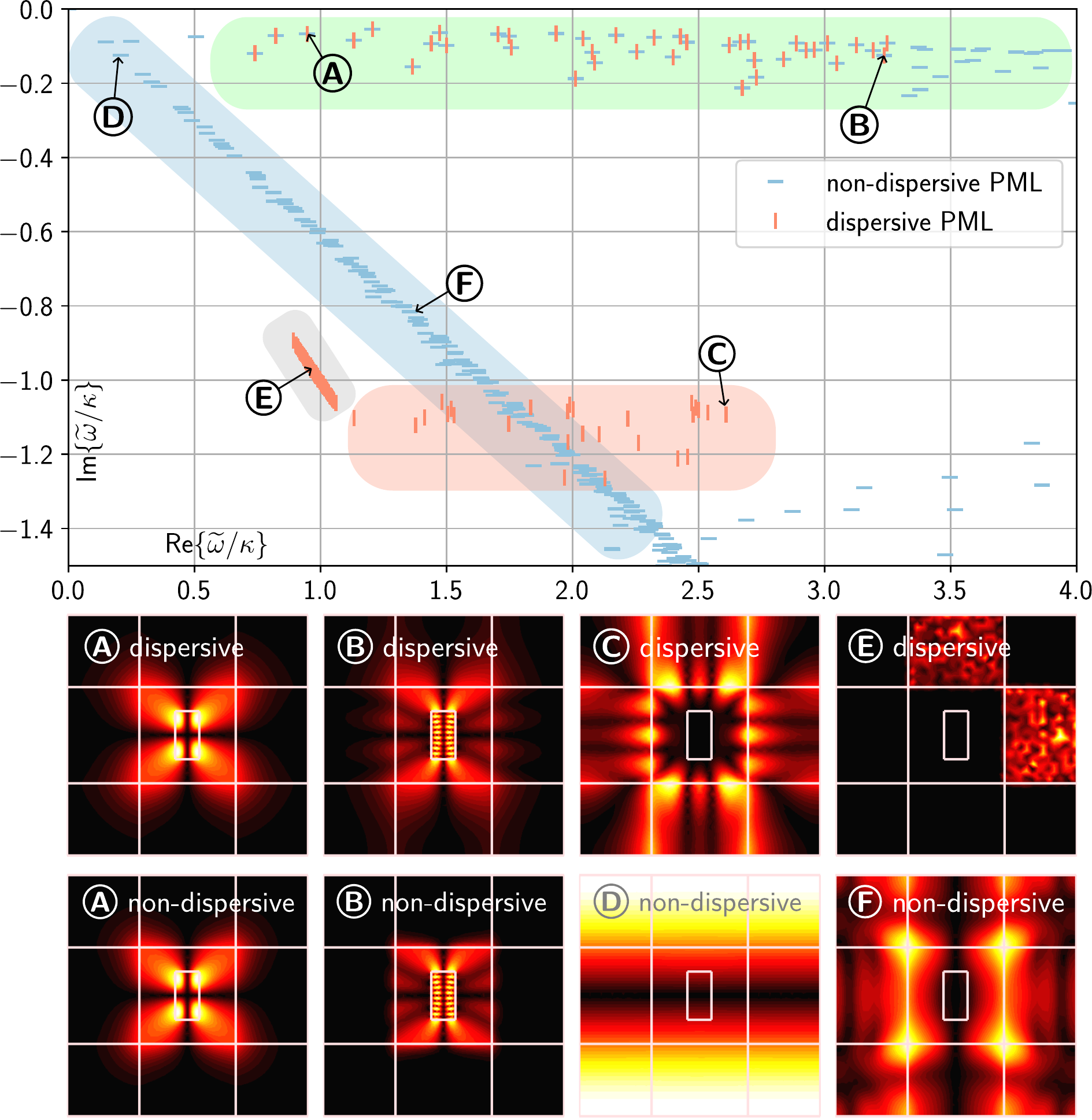}
  \caption{(Top right) Spectrum in the complex plane of eigenfrequencies (normalize by $\kappa=\pi c/d$) with non-dispersive PMLs (blue crosses) and with dispersive PMLs (orange pluses). The gray inset exhibits a zoom of a spurious accumulation region obtained in the dispersive case. The bottom field color maps represent the square norm of the eigenfields corresponding to the eigenfrequencies tagged by capital letters in the top right plot.}
  \label{fig:twodpmlsres} 
\end{figure}
\subsubsection{Eigenvalue problems and numerical results}
The relative permittivity and permeability tensors of the non-dispersive PMLs considered hereafter are obtained by considering $\redom=1$ in Eq.~\ref{eq:defpml}. 
Thus, the parametrization of PMLs (tensor fields $L_{xx}$, etc) becomes independent of $\redom$. 
The QNMs are then classically obtained by solving the following linear generalized eigenvalue problem:
\begin{equation}
  \left|
    \begin{array}{l}
      \mbox{Find non-trivial pairs $(\redom,\se)\in\mathbb{C}\times\bm{\mathcal{H}^1}(\Omega,\grad)$ such that :} \\[2mm]
      \quad \forall w\in\bm{\mathcal{H}^1}(\Omega,\grad),\\[2mm]
      \quad \quad -\Galerkind{\left(\begin{bmatrix}L_{yy}(\bx)&0\\0&L_{xx}(\bx)\end{bmatrix}\grad\,e\right)}{\grad\,\overline{w}}{\Omega}
      +\dfrac{\redom^2}{c^2}                  \Galerkinscald{{\tensepsrzz}(\bx) e}{\overline{w}}{\Omega}  =0 \,.
    \end{array}
  \right.
  \label{eq:weak2Dpmlnodisp}
\end{equation}

In the dispersive case, QNMs are obtained by solving the following polynomial eigenvalue problem of the third order :
\begin{equation}
  \left|
    \begin{array}{lll}
      \multicolumn{3}{l}{\mbox{Find non-trivial pairs $(\redom,\se)\in\mathbb{C}\times\bm{\mathcal{H}^1}(\Omega,\grad)$ such that:}} \\[2mm]
      \multicolumn{3}{l}{\quad \forall w\in\bm{\mathcal{H}^1}(\Omega,\grad),} \\[2mm]
      \quad\quad \dfrac{\redom^2}{\eta}       \Galerkinscald{\partial_x e}{\partial_x \overline{w}}{\Ompml{x}} 
    & -\eta                         \Galerkinscald{\partial_y e}{\partial_y \overline{w}}{\Ompml{x}}
    & -\dfrac{\eta\,\redom^2}{c^2}  \Galerkinscald{e}{\overline{w}}{\Ompml{x}}\\[3mm]
     \quad\quad-\eta                         \Galerkinscald{\partial_x e}{\partial_x \overline{w}}{\Ompml{y}} 
    & +\dfrac{\redom^2}{\eta}       \Galerkinscald{\partial_y e}{\partial_y \overline{w}}{\Ompml{y}}
    & -\dfrac{\eta\,\redom^2}{c^2}  \Galerkinscald{e}{\overline{w}}{\Ompml{y}}\\[3mm]
      \quad\quad-\redom\Galerkind{\grad\,e}{\grad\,\overline{w}}{\Ompml{xy}} 
    & \multicolumn{2}{l}{+\dfrac{\redom\eta^2}{c^2}    \Galerkinscald{e}{\overline{w}}{\Ompml{xy}}}\\[3mm]
      \quad\quad-\redom \Galerkind{\grad\,e}{\grad\,\overline{w}}{\Omega_b\cup\Omega_s}
    & \multicolumn{2}{l}{+\dfrac{\redom^3}{c^2}\Galerkinscald{{\tensepsrzz}(\bx) e}{\overline{w}}{\Omega_b\cup\Omega_s}  =0} \,.
    \end{array}
  \right.
  \label{eq:weak2Dpmldisp}
\end{equation}
The eigenvalue spectrum (normalize by $\kappa=\pi c/d$) obtained in the non-dispersive (resp. dispersive) case is represented by blue crosses (resp. orange pluses) in the complex plane at the top right of Fig.~\ref{fig:twodpmlsres}. 

The first striking observation is that the QNMs of the scatterer can be accurately retrieved using dispersive PMLs, as can be observed from the matching markers lying in the greenish part of the complex plane. In the non-dispersive (resp. dispersive) case, eigenfrequencies lying in the blueish (resp. orangish) region of the complex plane correspond to PML modes, exhibiting a rotation (resp. translation) of the real axis exactly as in the 1D toy model of the previous section. 

Further analysis shows that the QNMs of the eigenpairs labelled \circled{A} and \circled{B} are identical in the scatterer and freespace portions. They differ in the PML region as expected. In the case \circled{A}, the damping is similar in both dispersive and non-dispersive cases since $\omeig_{\mathsf{A}}/\kappa\approx 1$. However, in the case \circled{B} at $\omeig_{\mathsf{B}}/\kappa\approx 4$, the spatial damping of the QNM is expected to be stronger in the non-dispersive case than that of the dispersive one, which is confirmed by the eigenfields \circled{B}. Finally, the damping obtained in cases \circled{A} and \circled{B} are similar in the dispersive case, and we retrieve the desired feature of the dispersive PML as already shown in 1D.

The eigenfields \circled{C}-\circled{F} correspond to PMLs modes. Low frequency modes in the non-dispersive case, as mode \circled{D} at $\omeig_\mathsf{D}/\kappa\approx0.2-0.2i$, are similar to the fundamental modes of the bare cavity $\Omega$: The skin depth of the finite size PML is far too large. The mode \circled{C} is a PML mode in the dispersive case ; the eigenfield is pinched at each $\Ompml{xy}/\Ompml{x}$ and $\Ompml{xy}/\Ompml{y}$ interfaces and reaches the outer boundary of $\Omega$ thanks to a guiding property. In fact, those modes can be qualified of plasmonic modes since we are facing the PML tensor coefficients are sign changing at those interfaces (see Eq.~\ref{eq:defpml}). 

Finally, the eigenvalues within the greyed region of the complex plane in Fig.~\ref{fig:twodpmlsres} have corresponding eigenfields similar to the eigenpair labelled \circled{E}. The field inside the PMLs is clearly noisy, as if the mesh was not refined enough to capture the oscillations of the eigenfield. 

\subsubsection{The 2D naive dispersive PML failure}
\begin{figure}[h] 
  \centering 
  \includegraphics[draft=\flagdraftfig,width=.98\textwidth]{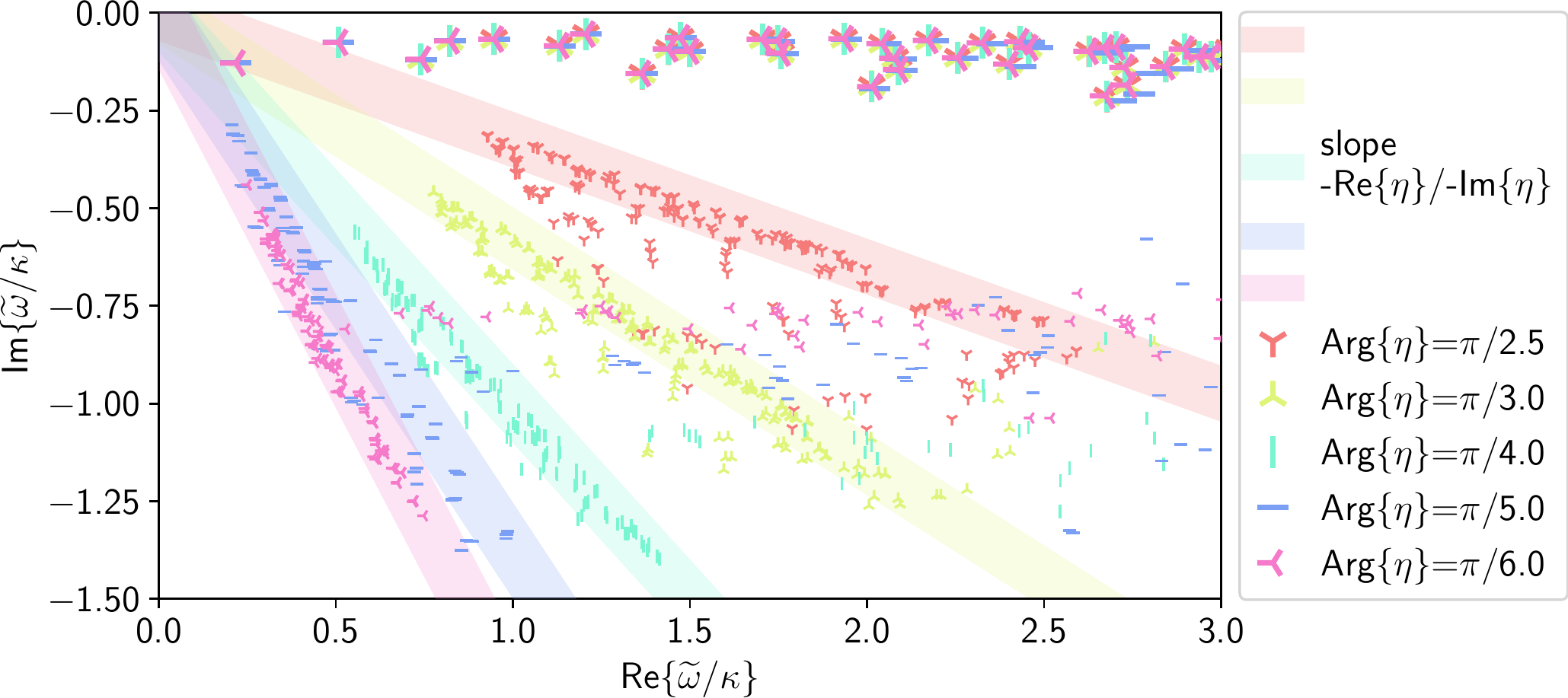}
  \caption{Spectrum of eigenfrequenies in the complex plane of the structure shown in Fig.~\ref{fig:twodpmls} computed with dispersive PMLs using various values of the stretch parameter $\eta$. The large lines have slopes $-\mathrm{Re}\{\eta\}/\mathrm{Im}\{\eta\}$ and indicate the unwanted branch parametrized by Eq.~\ref{eq:newbranch}.
  }
  \label{fig:spurious_slopes}
\end{figure}

Consider for instance a $b_x\times b_y$ square Cartesian dispersive PML medium adapted to the $x$ direction bounded by a homogeneous Dirichlet condition. Its permittivity and permeability are give in Eq.~\ref{eq:defpml}. The dispersion relation in such a medium becomes:

  \begin{equation}
    \begin{tikzpicture}[scale=1,baseline=(current bounding box.center)]
      \def \bx{1};
      \def \by{1.5};
      \def \myscale{0.7};
      \def \eps{0.05};
      \filldraw[fill=cyan!30, draw=red, thick] (0,0) rectangle (\bx,\by);
      \node[scale=1.5*\myscale] at (0.55*\bx,0.4*\by)  {$\Ompml{x}$};
      \draw[color=black, <->, >=latex] (0,\by-2*\eps) -- (\bx,\by-2*\eps) node[midway,below,scale=\myscale]{$b_x$};
      \draw[color=black, <->, >=latex] (-2*\eps,0) -- (-2*\eps, \by) node[midway,left,scale=\myscale]{$b_y$};
      \node[scale=\myscale,color=red] at (1.5*\bx,0.1*\by) { $\left.e\right|_{\partial\Omega}=0$ };
    \end{tikzpicture}
    \label{eq:disprel}
    \hspace*{1cm}\frac{k_x^2}{\mu_r^{yy}\,\varepsilon_r^{zz}} + \frac{k_y^2}{\mu_r^{xx}\,\varepsilon_r^{zz}}=\frac{\redom^2}{c^2}
  \end{equation}
which reduces to : 
\begin{equation}
  \label{eq:disprel2}
  k_x^2 + \frac{\eta^2\,k_y^2}{\redom^2}=\frac{\eta^2}{c^2}
\end{equation}
in a dispersive PML adapted to the $x$ direction. Given any $(n,m)$ positive integers, the eigenmodes classically write as $\sin(nk_x x)\sin(mk_y y)$ with $k_x=n\pi/d_x$ and $k_y=m\pi/b_y$, which leads to:
\begin{equation}\label{eq:newbranch}
  \redom=\sqrt{\frac{\eta^2k_y^2}{\eta/c^2-k_x^2}}\xrightarrow[n\to +\infty]{}\pm i \frac{m}{n}\frac{b_x}{b_y}\eta .
\end{equation}
At a fixed large $n$ value grows to infinity, it becomes obvious that increasing the value $m$ will gradually densely populate the line with slope $-\mathrm{Re}\{\eta\}/\mathrm{Im}\{\eta\}$. Even though the PML media along $x$ have more intricate boundary conditions in connection with the domain of interest and other PML, the same mechanism applies. A numerical experiment presented in Fig.~\ref{fig:spurious_slopes} shows the problematic lines (large lines in light colors) and the problematic eigenvalues with possibly infinite spatial frequencies along $y$ for various values of $\eta$.

\subsection{2D graded-indexed dispersive PMLs}
In the previous sections, we have shown that the dispersive PML causes a large number of modes with noisy fields inside the PML layer. In this section, we consider a parabolically damped graded-indexed dispersive Perfectly Matched Layers (PMLs) which can alleviate the issue \cite{gras2020nonuniqueness}. The PML relative permittivity and permeability tensors are given by $\boldsymbol{\mu}^{PML}_r(\bx)/\mu_b=\boldsymbol{\varepsilon}^{PML}_r(\boldsymbol{x})/\varepsilon_b=(-i\redom +\boldsymbol{\mathrm{T}}_{231})(-i\redom +\boldsymbol{\mathrm{T}}_{312})\left(-i\redom +\boldsymbol{\mathrm{T}}_{123}\right)^{-1}(-i\redom)^{-1}$ with $\boldsymbol{\mathrm{T}}_{ijk}=\mathrm{Diag}[\sigma_i \quad \sigma_j \quad \sigma_k]$ and damping coefficients $\sigma_i$ given by

\begin{equation}
\begin{cases}
\sigma_1 = m(|x| - |x_b|)^2, \quad \sigma_2 = \sigma_3 = 0 & \text{in } \Omega_x^{\mathrm{PML}} \\
\sigma_2 = m(|y| - |y_b|)^2, \quad \sigma_1 = \sigma_3 = 0 & \text{in } \Omega_y^{\mathrm{PML}} \\
\sigma_1 = m(|x| - |x_b|)^2, \quad \sigma_2 = m(|y| - |y_b|)^2, \quad \sigma_3 = 0 & \text{in } \Omega_{xy}^{\mathrm{PML}}
\end{cases}
\label{eq:graded}
\end{equation}

In Eq.~\ref{eq:graded}, $m$ is a real positive number that adjusts/controls the PML absorption coefficient (or damping rate), and $x_b$, $y_b$ denote the PML inner boundary. The PML has a graded index with progressively varying coefficients, the permittivity and permeability being the same as the background medium ($\varepsilon_b=1$, $\mu_b=1$) at the inner PML boundary. Perfectly electric conditions are applied to the outer border of the computational domain $\Omega$. Note that, in contrast to the dispersive PML discussed in previous sections, the permittivity and permeability satisfy the causality property $\boldsymbol{\varepsilon}^{PML*}_r(\redom)=\boldsymbol{\varepsilon}^{PML}_r(-\redom^*)$ and $\boldsymbol{\mu}^{PML*}_r(\redom)=\boldsymbol{\mu}^{PML}_r(-\redom^*)$. This has an advantage, as it allows one to deduce the QNMs with $\operatorname{Re}(\redom)<0$ from the QNMs with $\operatorname{Re}(\redom)>0$:  $e(\redom)=e^*(-\redom^*)$ \cite{sauvan2022normalization}.

\begin{figure}[h!]
  \centering
  \includegraphics[width=.9\textwidth]{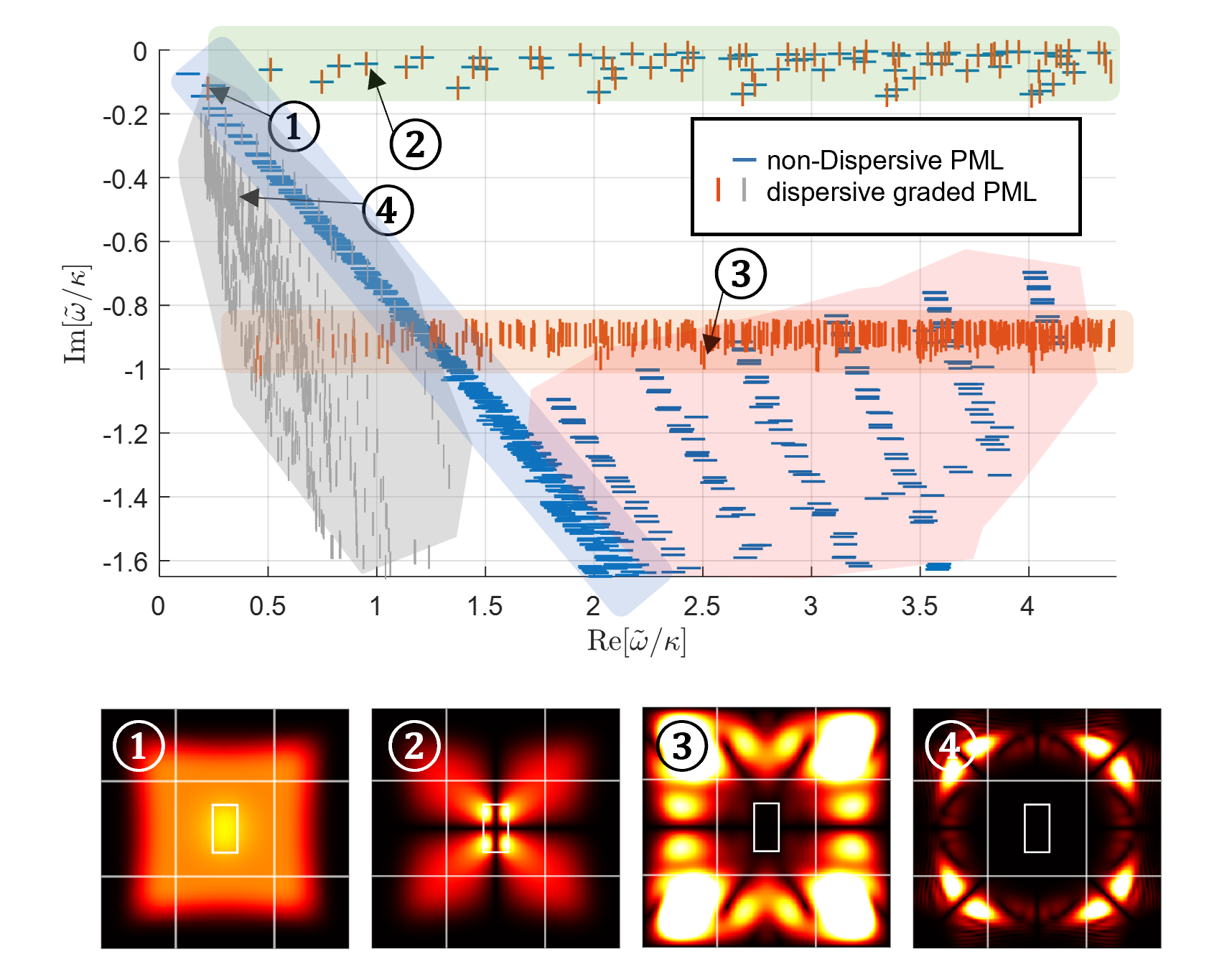}
  \caption{(Top) Comparison of the eigenmode spectra (normalize by $\kappa=\pi c/d$) computed with non-dispersive (blue horizontal ticks) and graded dispersive (orange vertical ticks) PMLs. (Bottom) The bottom field color maps represent the norm of the eigenfields (electric field z component) corresponding to the eigenfrequencies tagged by numbers in the top plot. }
  \label{fig:cpx 2}
\end{figure}

For the following numerical results, the QNMs are obtained by solving a polynomial eigenvalue problem:

\begin{equation}
  \left|
    \begin{array}{l} 
      \multicolumn{1}{l}{\mbox{  Find $(\redom,\se,P,\mathbf{M})\in\mathbb{C}\times\bm{\mathcal{H}^1}(\Omega,\grad)\times\bm{\mathcal{H}^1}(\Ompml{},\grad)\times\bm{\mathcal{H}^1}(\Ompml{},\curl)$}} \\
      \multicolumn{1}{l}{\mbox{ such that $\forall (w,p,\mathbf{m})\in\bm{\mathcal{H}^1}(\Omega,\grad)\times\bm{\mathcal{H}^1}(\Ompml{},\grad)\times\bm{\mathcal{H}^1}(\Ompml{},\curl)$:}} \\[2mm]
      \quad \Galerkinscald{\grad\,e\cdot}{\grad \overline{w}}{\Omega} 
      -\dfrac{\redom^2}{c^2}\Galerkinscald{\tensepsrzz e}{\overline{w}}{\Omega}
      \\[2mm]
      -\dfrac{\redom^2}{c^2}\Galerkinscald{\varepsilon_b P}{\overline{w}}{\Ompml{}}
      +\Galerkinscald{M_x\partial_y e -M_y\partial_x e}{}{\Ompml{}} =0
       \\[2mm]
       \quad \Galerkinscald{(\redom^2+i\redom \sigma_3)}{(e+P)\overline{p}}{\Ompml{}} 
       +\Galerkinscald{(i\redom-\sigma_1)(i\redom-\sigma_2)}{e\overline{p}}{\Ompml{}} 
       =0 
      \\[2mm]
      \quad \Galerkinscald{(-i\redom+\sigma_1)^2}{(\partial_y e)\overline{m}_x}{\Ompml{}}
      + \Galerkinscald{(-i\redom+\sigma_2)^2}{(-\partial_x e)\overline{m}_y}{\Ompml{}} 
      \\[2mm]
      -\Galerkinscald{(i\redom-\sigma_2)(i\redom-\sigma_1)}{[(M_x+\partial_ye)\overline{m}_x+(M_y-\partial_xe)\overline{m}_y]}{\Ompml{}}=0, 
       \\[2mm]
      
    \end{array}
  \right.
  \label{eq:weak2Dpmldisp}
\end{equation}
where $\Ompml{}:=\Ompml{x}\cup\Ompml{y}\cup\Ompml{xy}$ , and $P=(\varepsilon^{\text{\tiny PML}}_{rzz}-\varepsilon_b)e/\varepsilon_b$ and $\mathbf{M}=([\boldsymbol{\mu}^{\text{\tiny PML}}_r]^{-1}-1)\,\curl{}[e\mathbf{e_z}]$ are auxiliary fields.
We consider the same system as Sec.~\ref{part:fem2Ddisppml}. The computed eigenvalue spectrum (normalize by $\kappa = \frac{\pi c}{d}$) is shown with vertical ticks in Fig.~\ref{fig:cpx 2}. For comparison, we also show the spectrum for the case where a non-dispersive constant PML is used, as shown with the horizontal bars. The non-dispersive PML has a Polynomial coordinate stretching and can be viewed as a constant absorbing material as that described in Sec.~\ref{part:fem2Ddisppml}. The models for computing the results in this section can be downloaded in the latest version of the freeware MAN ~\cite{wu2023modal}.

The spectrum with the graded PML encompasses three different types of modes: QNMs with low $\text{Im}(\tilde{\omega})$, e.g. modes \circled{1} and \circled{2}; PML modes, which are distributed in a straight line parallel to the real axis, e.g. mode \circled{3}; and a myriad of PML modes, which are randomly distributed in the grey region, e.g. mode \circled{4}. The eigenfrequencies in the green region, e.g mode \circled{2}, are the same for the two PMLs, except for very low frequencies $\text{Re}(\tilde{\omega}/\kappa)<0.5$. In the non-dispersive case, the low frequency QNMs are contaminated by the rotated PML modes: they cannot be computed since their eigenfrequencies and field distributions depend on the PML damping factors and thickness. In the dispersive case, conversely, the QNMs with low frequency, e.g. mode \circled{1}, are independent of the PML properties: the computed eigenvectors are real QNMs.

The spectrum of the dispersive case contains two different types of PML modes. The first type (orange region), e.g. mode \circled{3}, are nearly aligned on a horizontal line parallel to the real axis. Like the PML modes of the non-dispersive case, they originate from the continuum spectrum. The second type (grey region), e.g. mode \circled{4}, are distributed in the low frequency region. Similar to the mode \circled{F} discussed in previous sections, these modes are also caused by the large spatial frequency waves inside PML layers. Thanks to the graded feature of the PML layer, the electric fields of these PMLs vary smoothly.

Finally, it is worth mentioning that, compared to the non-dispersive PML, the dispersive PML efficiently prevents the occurrence of Trefethen pseudomodes \cite{trefethen2006nonhermitian}. As shown in Fig. \ref{fig:cpx 2}, the Trefethen pseudomodes (modes in the reddish region) occur in the high-frequency region, when the limited mesh density fails to preserve the PML properties. The permittivity of the dispersive PML layer decreases as the frequency increases, providing an advantage over non-dispersive PML in preventing the occurrence of these non-physical modes. The current mesh, which was unable to capture the properties of non-dispersive PML, is able to accurately capture the behavior of dispersive PML. As can be seen in Fig. \ref{fig:cpx 2}, no Trefethen pseudomode occurs when the dispersive PML is implemented.

\subsection{Model expansion by implementing dispersive PML}
In this section, we study the accuracy of the reconstruction using a modal expansion by implementing dispersive and non-dispersive PMLs discussed in the previous section. We consider a plane wave incident on the same structure, given by $ e_i (\bold{x},\omsrc)=\exp(i\bold{k}\cdot \bold{x})$ with $\bold{k}=\frac{\omsrc}{c}\bold{e}_y$ and $\omsrc\in\mathbb{R}^+$. The frequency $\omsrc$ is the real driving frequency of the incident wave. We utilize the modal-expansion approach to reconstruct the scattered field $ e_s=\sum_{m}\alpha_m(\omsrc)\tilde{e}_m $, where $\alpha_m(\omsrc)=\frac{\omsrc \delta \varepsilon}{(\tilde{\omega}_m-\omsrc)}\int_{\Omega_s}\tilde{e}_m(\bold{x}) e_i(\bold{x},\omsrc)\,\mathrm{d}\Omega$ is the QNM excitation coefficient \cite{bai2013efficient}. $e_m$ is the normalized QNM field \cite{sauvan2013theory} for the mode with eigenfrequency of $\tilde{\omega}_m$ and $\delta\varepsilon = (\varepsilon_r-1)\varepsilon_0$ is the difference between the permittivity of the resonator and that of the background. With the knowledge of the scattered field, we further compute the extinction cross section using the formula $\sigma_{ext} = \frac{\mu_0c\omsrc}{|e_i|^2}\int_{\Omega_s} \mathrm{Im}[\delta\varepsilon(e_s+e_i)e_i^*]\,\mathrm{d}\Omega$.

\begin{figure}[h!]
  \centering
  \includegraphics[width=0.98\textwidth]{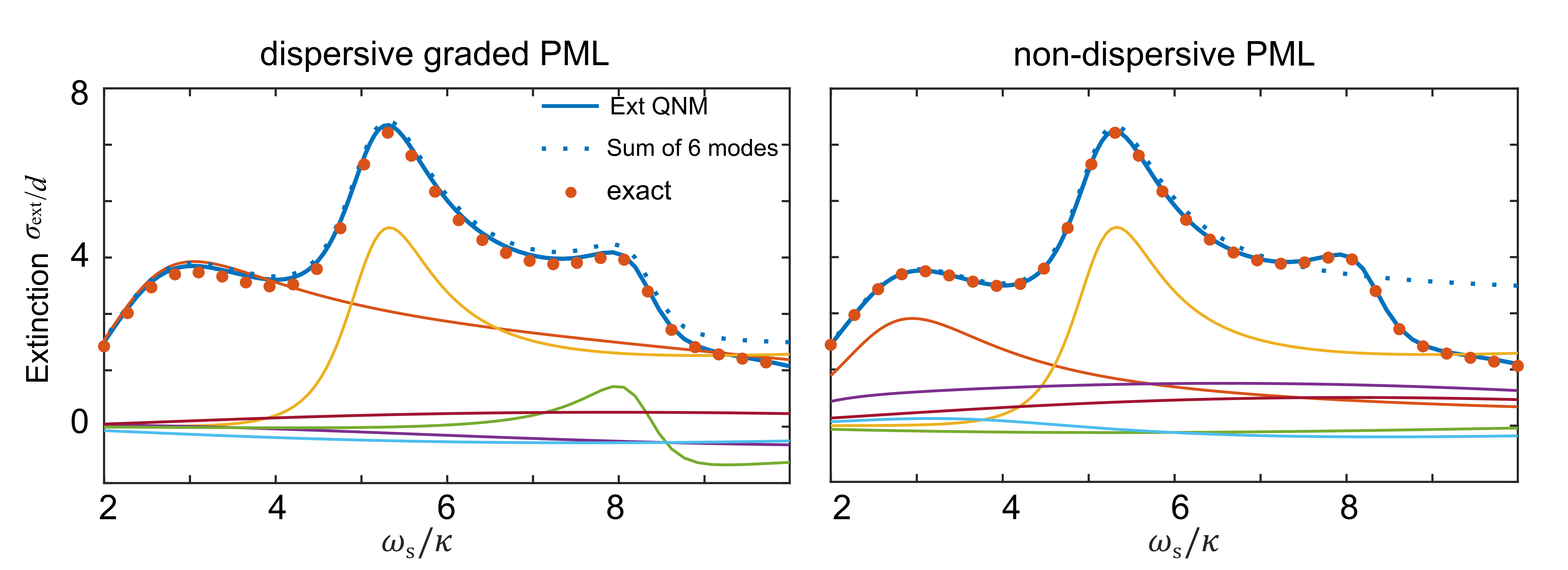}
  \caption{(Top) Extinction cross-section spectra reconstructed with 6 dominant QNMs only (blue dotted curves) and 200 modes including QNMs and TDMs (blue solid curves) are compared with reference ``exact'' data (circles) computed by directly solving the Maxwell equation at real frequency. The contributions of every 6 dominant modes are shown with the colored thin solid lines.}
  \label{fig:cpx 3}
\end{figure}

Figure \ref{fig:cpx 3} compares the results obtained by implementing dispersive and non-dispersive PMLs. To study the accuracy of the reconstruction, we also show the extinction computed by directly solving the Maxwell equation at real frequency (circles). 

Our analysis focuses on two types of convergence rate: asymptotic rates obtained for a large number of modes retained in the expansion, and initial rates obtained when only a few dominant modes that govern the physics of the system are retained in the expansion.

The blue dotted lines in Fig. \ref{fig:cpx 3} show the reconstructed extinction when only the six dominant modes are retained in the expansion. We first note that a good agreement is achieved over almost the entire spectral range for the dispersive-PML, except at high frequencies for which the absence of high-order QNMs leads to slight inaccuracies. Conversely, the non-dispersive PML yields excellent agreement at low frequencies but provides significant errors at higher frequencies. We also note that the extinction for low frequencies $\omsrc\in[2\kappa,4\kappa]$  is well reconstructed with a single QNM for the dispersive case, whereas several modes are required for accuracy for the non-dispersive case. As we checked numerically, these observations are due to the contamination of the low-frequency QNMs by the PML modes for the non-dispersive case. Owing to the contamination, more modes require to be retained in the expansion to capture the main features of the spectrum.

We further compare the reconstruction accuracies of the two approaches when 200 modes are retained in the expansion (blue solid curves in Fig. \ref{fig:cpx 3}). Unlike the previous case, a higher accuracy is achieved with the non-dispersive PML. The QNM reconstruction and the direct results exactly overlap; the spectrally averaged error $\left\langle|\sigma_{ext}^{(direct)}-\sigma_{ext}^{(QNM)}  |/\sigma_{ext}^{(direct)}\right\rangle$ is $~10^{-3}$ and can be further reduced by using a finer discretization. Conversely, for the dispersive case, there is a visible discrepancy of ~4\%. We checked that further increasing the number of modes does not significantly reduce the errors. We believe that the bad asymptotic convergence rate for the dispersive case is due to the existence of numerous PML modes with large spatial frequency (such as mode \circled{4} in Fig. \ref{fig:cpx 3}) in the low spectral range.

\color{mycolblack}
\section{Absorbing boundary conditions}\label{sec:dtn}

\begin{figure}[h!]
  \centering \def\svgwidth{.7\textwidth}
  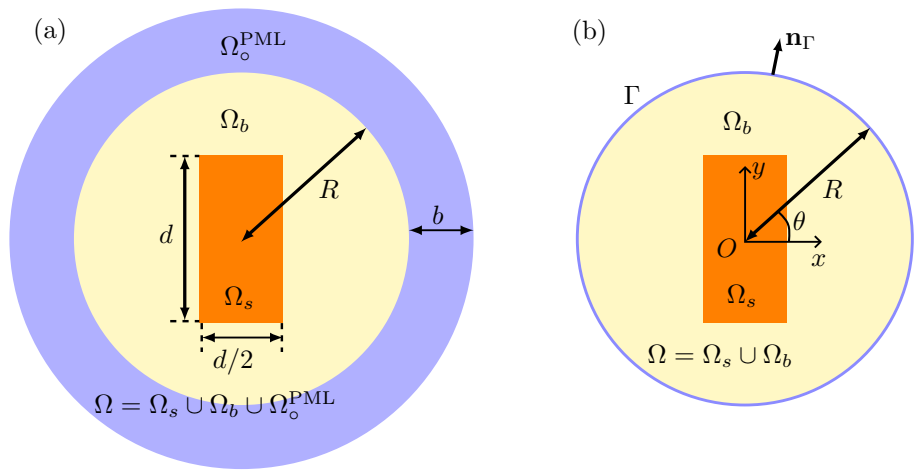
  \caption{Schematics of the 2D open structures.}
  \label{fig:schemedtn}
\end{figure}

The exact boundary condition at infinity for open 2D scalar problems is the 2D scalar Sommerfeld condition $\lim_{r\rightarrow\infty}\sqrt{r}[\partial_r u-i\redom u/c]=0$. This Sommerfeld condition is obviously frequency dependent and the limit as $r$ grow to infinity is of course unreachable with the FEM bounded mesh. When considering a bounded domain, an exact frequency-dependent operator $\mathcal{D}_\redom$ mimicking the outgoing wave condition can still be defined and connects the Dirichlet data (related to the unknown field) to the Neumann ones (related to the exterior derivative of the unknown field) -- this is a Dirichlet-to-Neumann map. For a circular exterior boundary as shown in the amagnetic problem described in Fig.~\ref{fig:schemedtn}(b), the QNM eigenvalue problem formally reads \cite{lenoir1992variational,araujo2018efficient,ARAUJOC2021110024}:
\begin{equation}
\begin{aligned}
  &-{c^2}\tensepsrzz(\x)^{-1}\,\div\left[\grad\,\se \right]={\redom^2}\se\;\mbox{ with }\;\grad\,e\cdot\bn_{\Gamma} = \mathcal{D}_\redom\,e\;\mbox{ on $\Gamma$, where:}\\
  & \mathcal{D}_\redom\,e(\theta) = \frac{\redom}{2\pi c} \sum_{n=-\infty}^{+\infty}
  \frac{H^{(1)'}_n(R\redom/c)}{H^{(1)}_n(R\redom/c)}
  \int_0^{2\pi}{e}(R,\varphi)\,\exp\left[in(\theta-\varphi)\right]\,\mathrm{d}\varphi  
\end{aligned}
\end{equation}

Besides being frequency-dispersive, $\mathcal{D}_\redom$ is also non-local through the Fourier-Bessel integrals in the above expression. Note that a modal boundary condition has been shown in Refs.~\cite{araujo2018efficient,ARAUJOC2021110024}. Finding local approximations of $\mathcal{D}_\redom$ has been the subject of intensive research \cite{antoine1999bayliss,modave2020corner}. We show in this section the numerical behavior of the first and second order approximations of $\mathcal{D}_\redom$ upon the determination of the QNMs. The first order is simply known as the $1^{\text{st}}$ order ABC (ABC1)), or Mur ABC, and the second one as the \textit{second order Bayliss-Turkel} (BT2) condition \cite{bayliss1980radiation,antoine1999bayliss} -- both can be treated as a generalized Robin condition, as detailed in the next two paragraphs. This condition is applied to the same rectangular object as in Sec.~\ref{part:fem2Ddisppml} standing in a freespace truncated with an ABC1 or BT2 condition at a distance $R$ from the center of the object (see Fig.~\ref{fig:schemedtn}(a)) or with a cylindrical PML as a reference (see Fig.~\ref{fig:schemedtn}(b)).

\subsection{Order 1: First order ABC (ABC1)}
The first order is known as the $1^{\text{st}}$ order ABC (ABC1) and reads as a finite-distance Sommerfeld condition: $\mathcal{D}_\redom\,e\approx(i\redom/c)e$. It can be easily treated as a mere Robin condition on $\Gamma$ and the corresponding variational formulation unveils a quadratic EVP:
\begin{equation}\label{eq:pepABC}
  \left|
  \begin{aligned}
    &  \mbox{Find $(\redom,\se)\in\mathbb{C}\times\bm{\mathcal{H}^1}(\Omega,\grad)$ such that:} \\
    & \quad \forall w\in\bm{\mathcal{H}^1}(\Omega,\grad),\\
    & \quad\quad-\Galerkind{\grad\, \se}{\grad\,\overline{w}}{\Omega} + \frac{i\redom}{c}\Galerkinscalline{\se}{\overline{w}}{\Gamma} + \frac{\redom^2}{c^2}\Galerkinscal{{\epsr}_{zz}(\x)\,\se}{w}{\Omega}=0.
  \end{aligned}
  \right.
\end{equation}
Note that the only parameter ruling the behavior of the ABC1, the truncation radius $R$, does not appear in the above formulation. 
\subsection{Order 2: Second order Bayliss-Turkel (BT2) condition}
For a 2D circular domain of radius $R$, the classical second order Bayliss-Turkel condition BT2 \cite{bayliss1982boundary,antoine1999bayliss} consists in the following local approximation of $\mathcal{D}_\redom$ :
\begin{equation}
  \mathcal{D}_\redom\,e\approx \frac{i\redom}{c} e - \frac{1}{2R} e -\frac{1}{8R^2(i\redom/c-1/R)} e + \frac{R}{2Ri\redom/c-2}\grad_t^2 e = 0,
  \label{eq:BT}
\end{equation}
where $\grad_t$ is the tangential gradient on $\Gamma$. This approximation can, again, be treated under variational formulation as an enriched Robin condition which now unveils a rational EVP:
\begin{equation}\label{eq:nepBT}
  \left|
  \begin{aligned}
    & \mbox{Find $(\redom,\se)\in\mathbb{C}\times\bm{\mathcal{H}^1}(\Omega,\grad)$ such that:} \\
    & \quad\quad \forall w\in\bm{\mathcal{H}^1}(\Omega,\grad),\\
    & \quad\quad\quad\quad -\Galerkind{\grad\, \se}{\grad\, \overline{w}}{\Omega}+\mathcal{R}_1(\redom) \Galerkinscal{{\epsr}_{zz}(\x)\,\se}{w}{\Omega}\\
    & \quad\quad\quad\quad +\mathcal{R}_2(\redom) \Galerkinscal{\se}{w}{\partial\Omega}+\mathcal{R}_3(\redom) \Galerkindlin{\grad\, \se}{\grad\, \overline{w}}{\Gamma}=0,\\
    & \quad\quad\mbox{where $\mathcal{R}_1$,$\mathcal{R}_2$ and $\mathcal{R}_3$ are rational functions of $\redom$}:\\
    & \quad\quad\mbox{$\mathcal{R}_1(\redom)=\dfrac{\redom^2}{c^2}$,
                      $\mathcal{R}_2(\redom)=i\redom/c + \dfrac{3-4Ri\redom/c}{8R-8R^2i\redom/c}$,
                      $\mathcal{R}_3(\redom)=\dfrac{-R}{2Ri\redom/c-2}$.}
  \end{aligned}
  \right.
\end{equation}
Note that both ABC1 and BT2 conditions introduce complex terms making the problem non-Hermitian, which is a key to unveil the QNMs.

\begin{figure}[t]
  \centering
  \includegraphics[draft=\flagdraftfig,width=.99\textwidth]{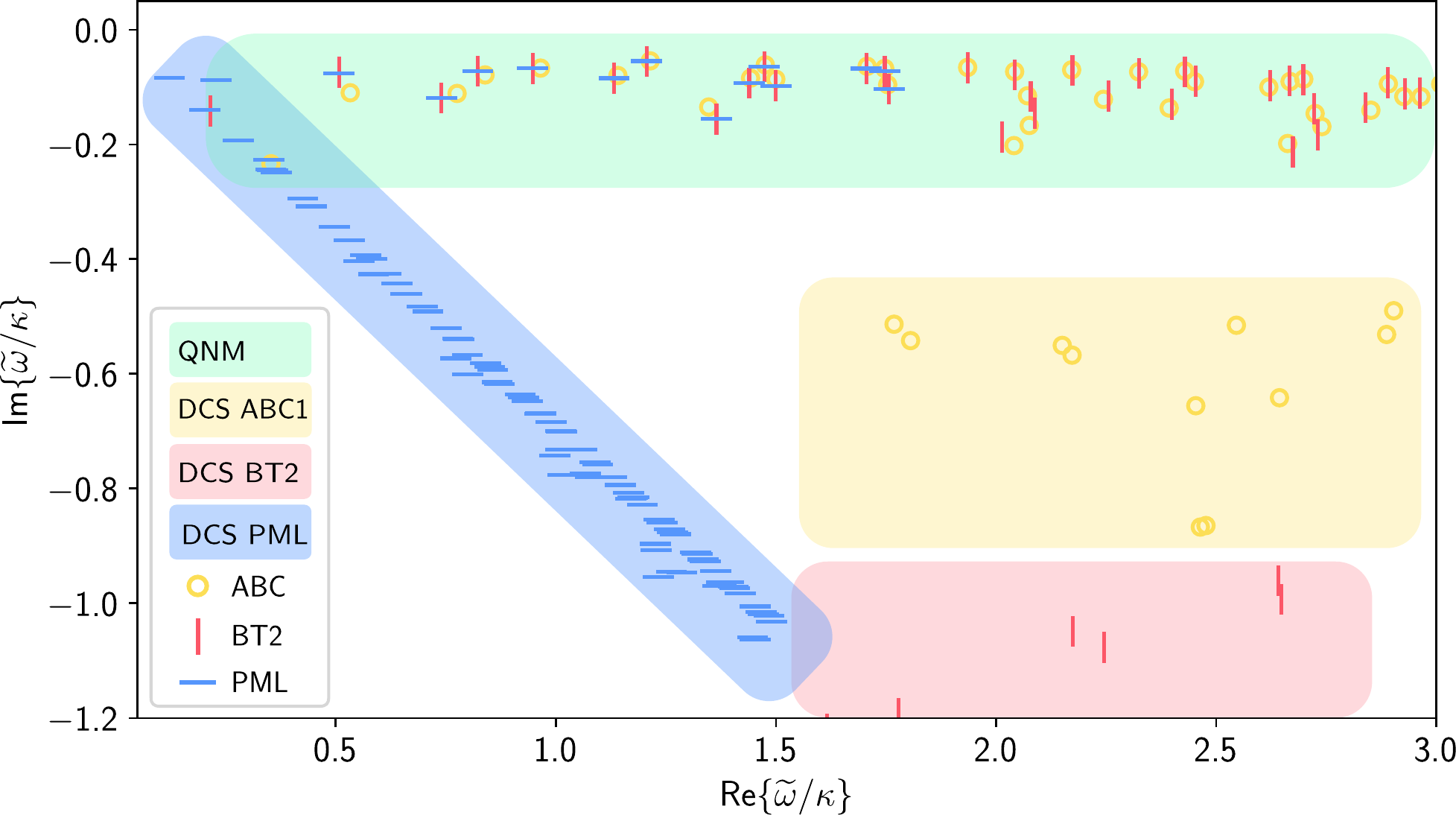}
  \caption{Spectrum in the complex plane of eigenfrequencies (normalize by $\kappa=\pi c/d$) computed with an ABC1 boundary condition (yellow circles), a BT2 boundary condition (red vertical dashes) and PMLs (blue horizontal dashes). The colored shaded regions highlight the QNMs of the scatterer (green area), the TDMs obtained with PMLs (blue area), the TDMs obtained with an ABC1 condition (yellow area) and a BT2 condition (red area).}
  \label{fig:eigs_dtn}
\end{figure}



\subsection{Comparison with QNMs obtained with a cylindrical PML} 

The QNMs and TDMs obtained for $R=d$ with the ABC1, BT2 boundary conditions and a cylindrical PMLs are depicted in Fig.~\ref{fig:eigs_dtn}.

\begin{figure}[h!]
  \centering
  \includegraphics[draft=\flagdraftfig,width=.99\textwidth]{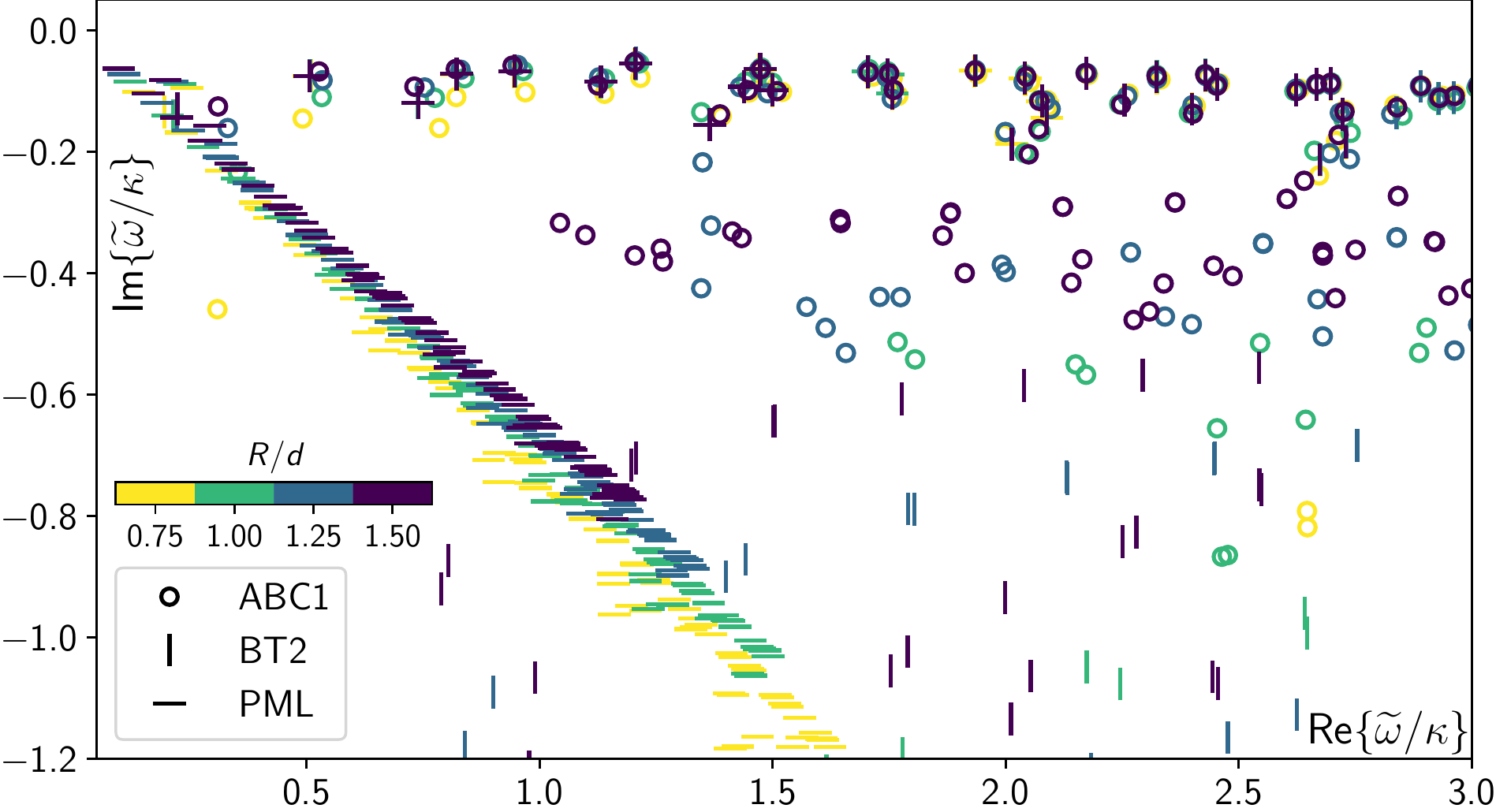}
  \caption{Spectrum in the complex plane of eigenfrequencies (normalize by $\kappa=\pi c/d$) computed with an ABC1 boundary condition (circles), a BT2 boundary condition (vertical dashes) and PMLs (horizontal dashes) for various values of the truncation radius $R$ given by the color-scale.}
  \label{fig:eigs_dtnR}
\end{figure}

The first striking observation concerns the greenish region close to the real axis corresponding to the QNMs. The QNMs found using a BT2 boundary condition  are in excellent agreement with those obtained with PMLS where the blue horizontal dashes and red vertical dashes form a plus sign. The agreement with QNMs found with the ABC1 boundary condition (yellow circles) is more rough, as could be expected, especially at low frequency where the apparent wavelength $\widetilde{\lambda}=\mathrm{Re}\{2\pi c/\widetilde{\omega}\} $ is large and thus the distance $R$ expressed in terms of $\widetilde{\lambda}$ is small: The approximate boundary condition is applied in the very near field at these lower eigenfrequencies. These observations are further confirmed when varying the truncation radius $R$ from $0.75\,d$ to $1.5\,d$, as shown in the complex frequency plane Fig.~\ref{fig:eigs_dtnR}. The data corresponding to the various radii are represented using a colormap. It is clear that the QNMs obtained with the PML and the BT2 boundary condition are independent of the truncation radius (all vertical and horizontal dashes corresponding to QNMs are superimposed). The low frequency QNMs obtained with the ABC1 condition move closer and closer to the PML/BT2 QNMs as $R$ increases.

The TDMs obtained for the ABC1 (yellowish region) and BT2 (reddish region) is radically different from the usual rotation in the complex plane obtained with PMLs (blueish region). Moreover, those TDMs are much less dense than that obtained with PMLs. In fact, in order to discover a number of matching QNMs  large enough for a fair comparison, the number of eigenvalues requested was set to 100 in the ABC1/BT2 case and 150 in the PML case. The ratio between the number of QNMs unveiled and the number of TDMs is clearly in favor of the ABC1/BT2 case, which is consistent with the fact that ABC1 and BT2 conditions are approximations of the $\mathcal{D}_\redom$ operator ; the approximation of the continuous spectrum may be less rich than when using PMLs, but \textit{in fine} leads to an accurate approximation.

\subsection{Modal expansion}
\begin{figure}[h!]
  \centering
  \includegraphics[draft=\flagdraftfig,width=.99\textwidth]{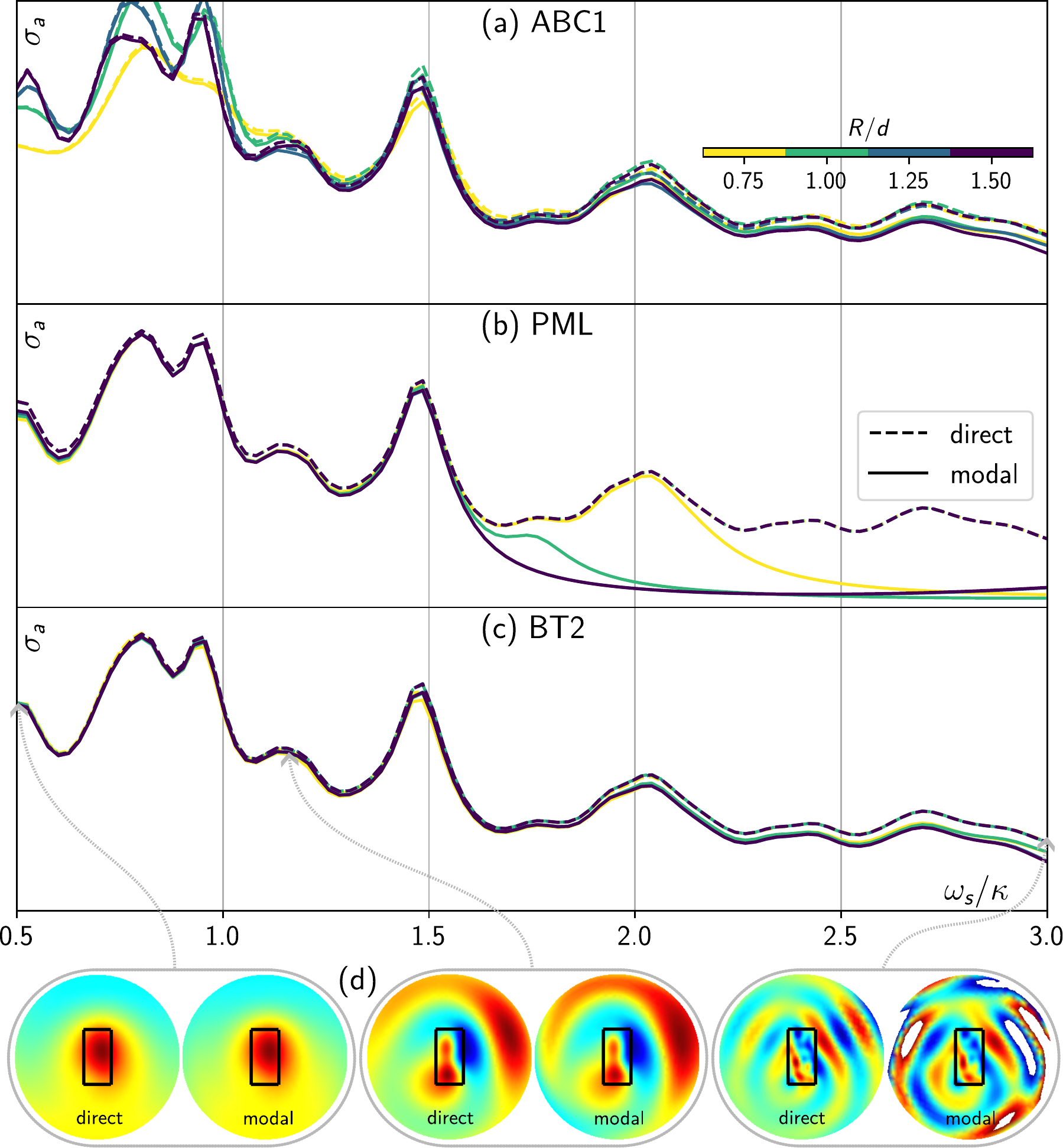}
  \caption{Absorption cross-section spectra (for an incident plane wave) computed using a direct (dashed lines) and a modal (plane lines) approach with (a) an ABC1 boundary condition, (b) PMLs and (c) a BT2 boundary condition for various values of the truncation radius $R$ given by the color-scale. Color-maps (d) show the real part of the scattered field for selected real driving normalized frequencies $\omsrc/\kappa$ and for the two modal/direct approaches using a BT2 boundary condition.}
  \label{fig:expansion_dtn}
\end{figure}

Finally, we tackle a direct problem using a QNMs expansion. We consider the same rectangular structure enlighten by a plane wave 
$$e_i(\bx,\omsrc) = \exp[i\bk\cdot\bx] \mbox{ with } \bk=\omsrc/c[\cos(\theta) ,\sin(\theta)]\mbox{ and } \omsrc\in\mathbb{R}^+.$$
The frequency $\omsrc$ is the (real) driving frequency of the problem, and should not to be mistaken with the (complex) eigenfrequencies $\redom$. The scattered field formulation of problem classically amounts to find the scattered field $e_s$ defined as the difference between the total field and the incident field solution of:
\begin{equation}\label{eq:directpb}
  \div[\tensmur^{-1}\grad\,e_s]+\frac{{\omsrc}^2}{c^2}\,\tensepsrzz\,e_s = \frac{{\omsrc}^2}{c^2}(\varepsilon_r^{bg}-\varepsilon_r^{scat}) e_i\,\mathsf{I_{\Omega_{s}}},
\end{equation}
where $e_s$ satisfies an outgoing wave condition. Once the eigenvalue problem solved, a QNM expansion of the scattering problem is performed (see Appendix). The absorption cross-section, proportional to $\sigma_a=\int_{\Omega_{s}}|e_s+e_i|^2\,\mathrm{d} S$, is computed using both approaches and the results are shown in Fig.~\ref{fig:expansion_dtn} for various truncation radii. The expansions allow retrieving quantitatively the results of the direct approaches with matching approximations $\mathcal{D}_\redom$ (\textit{i.e.} in Fig.~\ref{fig:expansion_dtn}(a), an ABC1 condition is used both in the direct and modal approaches).

For the ABC1 condition, we retrieve at low frequency $\omsrc$ the signature that low frequency QNMs were inaccurately computed and truncation radius dependent. For the non-dispersive PML case, the number of relevant QNMs in the frequency range considered $\omsrc\in[0.5,3]$ is too low to retrieve the full absorption cross-section spectrum. At constant number of eigenvalues (150), smaller radii $R$ allow unveiling more QNMs (see the plain yellow and dark blue lines in Fig.~\ref{fig:expansion_dtn}(b)). Other discrepancies are observed at low frequencies in the PML case. Finally, as for the BT2 case, with 100 eigenvalues, the full range of real driving frequencies $[0.5,3]$ is retrieved, with excellent agreement in the lower part of the range, as can be observed as well on the field maps shown in Fig.~\ref{fig:expansion_dtn}(d) where Re$\{e_s\}$ is represented with the same color-scale for the direct and modal cases. At higher frequencies, more contributions from the TDMs are necessary to reach a better convergence.


\section{Conclusion}

In this paper, we answer the crucial question of whether the dispersive PMLs or ABCs offer advantages in terms of stability and convergence rate compared to non-dispersive PMLs. Dispersive PMLs show promising features since they allow to damp the QNMs independently of the real part of their eigenfrequency. This addresses the problem of low frequency QNMs located near the origin of the complex plane that are veiled by the rotated discrete branch of non-dispersive PMLs. Additionally, they effectively eliminate the occurrence of spurious modes in the high-frequency region. Finally, the discrete translated PML branch obtained with dispersive PMLs is much sparser than its non-dispersive counterpart, so that, with a constant number of PML modes computed, the number of unveiled QNMs is larger with a dispersive PML than with a non-dispersive one. 
However, dispersive PMLs cause an unexpected problem when the dimension of the system is larger than 1D. We have shown that PMLs behave as a hyperbolic material in 2D within certain frequency ranges, supporting modes with possibly infinitely large spatial frequencies, which results in a large number of TDMs. The eigenfrequencies of these new modes lie on new branch in the complex plane corresponding to a rotation of the real axis and discretized by a rational quantity due to a cavity effect in the direction orthogonal to the damping one.  Their high spatial frequencies makes them difficult to compute accurately, which prevents to obtain very accurate reconstructions with modal expansion. A direction of further research consists in finding other types of PMLs which avoid the large spatial frequency modes. 
Furthermore, we discuss the advantages of using ABC boundary conditions in QNM computation and modal expansion. Compared to non-dispersive PMLs, we demonstrate that ABCs yield fewer TDMs within a given spectral range, enabling accurate modal expansion with a reduced number of modes. However, it is important to note that unlike PMLs, ABCs represent only a first-order local approximation of the non-local boundary condition that perfectly emulates outgoing wave conditions. Therefore, the reconstructed results do not precisely match the optical response of the actual infinite system. We also observe that the second-order local approximation BT2 leads to a much better agreement with the infinite system. Further research will be carried out in this direction to extend this concept to higher orders \cite{modave2020corner} and to the 3D vector case.

\appendix

\section*{Appendix: Numerical models}
\begin{itemize}
	\item \textbf{\texttt{ONELAB for photonics}}: The expansion makes use of the method described in Ref.~\cite{zolla2018photonics,NICOLET2022104809} relying on the softwares Gmsh \cite{gmsh}, GetDP \cite{Dular-GetDP} and SLEPc \cite{slepc}. A ONELAB template model named \texttt{QNM\_expansion\_DtN.pro} has been released \cite{ONELAB:QNMexpansionDtN} and allows to reproduce the results in Sec.~\ref{sec:dtn}.
	\item \textbf{\texttt{MAN}}: The COMSOL models for computing the QNMs in Fig. \ref{fig:cpx 2} are included in the MANMODELS folder of the freeware MAN \cite{wu2023modal}. The name of the models are \texttt{QNMEig\_ 2DnondispPML.mph} and \texttt{QNMEig\_2DdispPML.mph}.
\end{itemize}







\section*{Acknowledgements}
This work has been partially supported by the Agence Nationale de la Recherche (ANR) funded projects PLANISSIMO (ANR-12-NANO-0003) and RESONANCE (ANR-16-CE24-0013).

The authors acknowledge Wei Yan for fruitful conversations. All the members of the ANR project RESONANCE are also acknowledged.

\bibliographystyle{ieeetr}


\end{document}